\documentclass{elsart}

\newcommand\ba{\begin{eqnarray}}
\newcommand\ea{\end{eqnarray}}
\newcommand\bc{\begin{center}}
\newcommand\ec{\end{center}}
\newcommand\ra{\rightarrow}
\def\nn{\nonumber}

\newcommand\p{\partial}

\def\i{\ensuremath{\mathrm{i}}}

\def\O{\ensuremath{\mathcal{O}}}

\newcommand{\A}{\ensuremath{\mathbb{A}}}
\newcommand{\Origin}{\ensuremath{\mathbb{O}}}
\newcommand{\X}{\ensuremath{\mathbb{X}}}
\newcommand{\Y}{\ensuremath{\mathbb{Y}}}
\def\S{\ensuremath{\mathbb{S}}}
\newcommand{\T}{\ensuremath{\mathbb{T}}}
\def\L{\ensuremath{\mathbb{L}}}

\def\newchi{\raise2.0pt\hbox{\ensuremath{\chi}}} 

\def\epsilon{\varepsilon}
\def\e{\mathrm{e}}
\def\calS{\mathcal{S}}

\usepackage[dvips]{graphicx}
\usepackage{amsfonts}
\usepackage{latexsym}

\begin{document}
\begin{frontmatter}

\title{Instabilities induced by a weak breaking of a strong spatial resonance}

\author{J.~H.~P.~Dawes\thanksref{lab1}},
\author{C.~M.~Postlethwaite}, and
\author{M.~R.~E.~Proctor}

\thanks[lab1]{\emph{Corresponding author}}

\address{Department of Applied Mathematics and Theoretical Physics,
Centre for Mathematical Sciences, University of Cambridge, Wilberforce
Road, Cambridge, CB3 0WA, UK}

\begin{abstract}
Through multiple-scales and symmetry arguments we derive a model set of
amplitude equations describing the interaction of two steady-state 
pattern-forming instabilities, in the case that the wavelengths of the
instabilities are nearly in the ratio $1:2$. In the case of exact $1:2$
resonance the amplitude equations are ODEs; here they are PDEs. 

We discuss the stability of spatially-periodic solutions to
long-wavelength disturbances. By including these modulational effects
we are able to explore
the relevance of the exact $1:2$ results to spatially-extended
physical systems for parameter values near to this codimension-two
bifurcation point. These new instabilities can be described in terms
of reduced `normal form' PDEs near various secondary codimension-two
points. The robust heteroclinic cycle in the ODEs is destabilised by
long-wavelength perturbations and a stable periodic orbit is generated
that lies close to the cycle. An analytic expression giving the
approximate period of this orbit is derived.

\end{abstract}

\begin{keyword}
pattern \sep 
symmetry \sep
mode interaction \sep
bifurcation \sep
heteroclinic cycle \\
PACS Codes: 47.20 (bifurcation theory) ; 47.54 (pattern formation).
\end{keyword}

\end{frontmatter}

\section{Introduction}

\label{sec:intro}

Pattern-forming instabilities occur in many physical problems, for
example Rayleigh-B\'enard convection, Faraday wave experiments and
directional solidification \cite{CH90}. In some of these situations
well-established governing equations are available which are
sufficiently simple to analyse. In others the situation is not so
clear cut, and reductions to model
equations are of great value. The derivation of model equations to
describe instabilities in these physical problems often provides a
clear and unified viewpoint, bringing out similarities in the
underlying mathematical structure.

In one spatially-extended dimension, the study of the stability of
spatially-periodic patterns that arise from a steady-state bifurcation
with continuous translational symmetry often
reduces to the investigation of an evolution equation, for example the
Ginzburg--Landau equation. Such equations,
derived by asymptotic methods rather than rigorous analysis (though
see recent work by Melbourne \cite{M99}) can both comprehend uniform
spatially-periodic patterns and describe their stability to
long-wavelength disturbances.

In the present paper we consider the case when 
two distinct instability mechanisms are present, for example in 
two-layer thermal convection \cite{PJ88}. 
Here the two instabilities will typically have different
preferred horizontal lengthscales given by the critical wavenumbers
corresponding to local minima in
the curves of marginal stability. Each instability separately can be
described by a `universal' model equation such as the Ginzburg--Landau
equation. However, the two instability mechanisms interact nonlinearly
if the instabilities occur for similar values of the control
parameters; this is a codimension-2 bifurcation. In such a situation
more complicated model equations are needed to describe the dynamics
in the region of parameter space near the codimension-2 point.
The form of these model equations will depend on the symmetry of the
problem, the steady or oscillatory nature of the instabilities, and on
the ratio of the critical wavenumbers.

When the critical wavenumber ratio is rational (say $p/q$ where $p$
and $q$ are coprime integers) we may restrict
attention to solutions which are periodic in the horizontal direction
and derive, in a mathematically rigorous fashion using a centre
manifold reduction, ODEs describing the
dynamics on the centre manifold at the point of instability. The
cases $p=1$, $q=1,2,3$ have been termed `strong' spatial resonances
since the nonlinear interaction terms generated appear at third order,
or below, in the resulting amplitude equations. 

When the ratio is irrational this centre-manifold approach cannot be
applied. Any restriction to spatially-periodic solutions will not be
able to capture important features of the dynamics.
In consequence the periodic solutions found may be unstable to
long-wavelength instabilities, and their subsequent evolution must
be described by PDEs rather than ODEs. In this case, there is currently no
rigorous mathematical formulation which leads to PDEs. At best the
equations used represent asymptotic approximations to the true
situation.
Nonetheless we shall adopt the PDE approach here, following the work
of (among others) Coullet \& Repaux \cite{CR87}, since it is natural
and has proved very productive in similar, though simpler, problems.

In this paper we discuss the interaction of two steady-state
instabilities with wavenumbers close to, but not exactly in, the ratio
$1:2$, using multiple-scales expansions in time and space. The case of
exact $1:2$ resonance was treated first by Dangelmayr \cite{D86} and
later by Jones \& Proctor \cite{JP87} and
by Proctor \& Jones \cite{PJ88} (hereafter referred to as Part 1) in
the context of two-layer thermal convection. Important results were
also obtained by Armbruster et al. \cite{AGH87} and
Julien \cite{J91}, and more recently by Porter \& Knobloch \cite{PK00}. 
In particular, \cite{J91} resolved several contradictions and
errors in the earlier papers mentioned above, and investigated the
dynamics of the ODEs away from the codimension-2 point; a full
description of the dynamics of the ODEs becomes very involved. Here we
extend this work and show that 
small deviations from the exact $1:2$ situation result in
additional long-wavelength instabilities of otherwise stable spatially
periodic patterns (`spatial quasiperiodicity'). 

The most unusual part of the dynamics of the $1:2$ resonance problem
is the occurrence of a structurally stable heteroclinic cycle. The
local information that we are able compute near the equilibria on the
cycle enables analysis of the stability of the cycle to
long-wavelength spatial disturbances.

The equations that we derive to capture these modulational
instabilities are nonlinear PDEs, and as such, a full analysis is a
daunting task. Indeed, even a complete classification of possible
solutions is far beyond the scope of this paper. This paper considers
those parts of the problem that can be treated analytically rather
than presenting a superabundance of numerical results. 
Our analysis explores the stability of spatially-periodic equilibria and
travelling waves, identifying those new instabilities that are due to
the inclusion of modulational effects. The dynamics near these new
instabilities can be described by `normal form' equations (simpler
PDEs whose structure is often prescribed by symmetry
requirements). Although the algebraic expressions may become lengthy, 
it is possible to reduce the original nonlinear PDEs to these `normal
forms' explicitly via adiabatic elimination. Such a reduction, although
not completely rigorously justified, can be extremely useful,
particularly near secondary codimension-2 points (intersections of
lines of codimension-1 bifurcations away from the initial bifurcation
point). The paper illustrates this
idea with two explicit detailed examples (the points $\L$ and $\X$ in
figure~\ref{mu1mu2_fig}); in neither case is the `normal form'
equation the Ginzburg--Landau equation. The dynamics near $\L$ have
been well-studied in the literature, but those near $\X$ are more
novel.

The original study of Part 1 was motivated by a particular two-layer
thermal convection problem, and we refer the interested reader to that
paper for detailed discussion of the physical
background. Wavenumber interactions in the ratio $1:2$ also occur in
two-dimensional thermal convection in a single layer, for example in
the asymptotic long-wavelength equation discussed by Cox \cite{C96},
and in non-Boussinesq convection as discussed by Mercader, Prat \&
Knobloch \cite{MPK02,PMK02}. These papers discuss situations that
have one instability mechanism; by looking for solutions that are
spatially-periodic with a given wavenumber $k$ they locate
codimension-two points $(R_c,k_c)$ where two periodic patterns
interact, with wavenumbers in the ratio $1:2$. Since, in these
problems, $R_c$ is always greater than the minimum value required to
drive convection in a formally infinite layer, the results are not
directly applicable to the infinite layer case. In contrast, this
paper (formally) attempts to analyse the infinite-layer situation in
the case where there are two distinct instability mechanisms occurring
for very similar values of the control parameters. The large
horizontal extent of the layer enables periodic patterns to be
destabilised by `sideband' instabilities, for example the Eckhaus
instability. We remark that a similar study, for `weak' resonances,
has been performed recently by Higuera, Riecke \& Silber
\cite{HRS03}. Their results, although very different in detail, have
points of similarity to those presented here, for example the
existence of solutions in the form of localised structures.


The paper is organised as follows. In section~\ref{sec:model} we
derive the model equations and make general remarks about the
dynamics. In section~\ref{sec:analysis} we briefly summarise the
dynamics of the model ODEs in the absence of the modulational
terms. Here and throughout the rest of the paper we use a single
combination of coefficients that were used in Part 1, for ease of
comparison. Sections~\ref{sec:pure_mode} and~\ref{sec:mixed_mode}
discuss in detail the new instabilities of spatially-periodic states
that occur when the modulational
terms are added to the model. In section~\ref{sec:cycle} we
discuss the stability of the robust heteroclinic cycle. Analytic
work shows the existence of a long-period periodic orbit lying close
to the cycle, in quantitative agreement with numerical
results. Section~\ref{stable_tw_sec} briefly highlights the
coexistence, over a substantial region of the parameter plane, of
stable non-modulated travelling waves and complex spatiotemporal
behaviour. We conclude in section~\ref{sec:conclusions}.


\section{Model equations near $1:2$ resonance}

\label{sec:model}

Consider a horizontal two-dimensional layer, or layers, of fluid in
the domain $(-\infty,\infty) \times [0,1]$, using co-ordinates
$(x,z)$, i.e. of finite vertical extent but extending to infinity in
the horizontal direction $x$. We suppose that there is an
$x$-independent state which may become linearly unstable to either of
two competing steady-state instabilities, with wavenumbers nearly
in the ratio $1:2$. By way of illustration, the analysis of Part 1 was
concerned with
a two-layer thermal convection problem where these two instabilities
corresponded to the onset of convection predominantly in either the
upper or the lower layer separately. The most interesting dynamics can
be captured by
the distinguished limit in which the deviation of the ratio of
the critical wavenumbers from the exact value $\frac{1}{2}$ is of the
order of
the square root of the deviation of the bifurcation parameter (in this
case the Rayleigh number $R$) from its critical value $R_c$. 
If we set $R-R_c=\epsilon^2$ then we may write
the critical wavenumbers (those associated with local minima in the
value of the critical Rayleigh number) as $k+\epsilon q$ and $2k$. A
suitable ansatz for small-amplitude solutions near the codimension-2
point where the conditions for instability coincide is then
\ba
u(x,z,t) & = & \epsilon \left[ A(X,T) f_1(z) \e^{\i x(1 + \epsilon q)} +
B(X,T) f_2(z) \e^{2\i x} + c.c. \right] \nn \\
& & + O(\epsilon^2), \label{ansatz}
\ea
where lengths have been rescaled so that $k=1$; $X=\epsilon x$ and
$T=\epsilon^2 t$ are long length and time
scales, $c.c.$ denotes complex conjugate, and the functions
$f_{1,2}(z)$ give the vertical structure of the
eigenfunction corresponding to each mode of instability. The
amplitudes $A(X,T)$, $B(X,T)$ are complex-valued.
The initial homogeneous state is symmetric under the Euclidean group $E(1)$
generated by the reflection
$m_x: x \ra -x$ and the translations $\tau_\delta: x
\ra x+\delta$. These symmetries induce
the following transformations on the amplitudes $A$ and $B$:
\ba
m_x: 		& x \ra -x :	& \qquad (A,B) \ra ( \bar{A} ,
\bar{B} ), \label{reflection_sym} \\
\tau_\delta: 	& \ \ x \ra x + \delta: & \qquad (A,B) \ra ( A \e^{\i ( 1 + \epsilon
q ) \delta}, B \e^{2 \i \delta} ). \label{translation_sym}
\ea
The combination $\bar{A}^2 B \e^{2\i qX}$ is found to be the
lowest-order translation-invariant combination that is not a product
of the usual terms $|A|^2$ and $|B|^2$.

The resulting amplitude equations (ignoring
terms of order higher than three in $A$, $B$ and $\p_X$) take the form
\ba
\dot{A} & = & A [\mu_1 - a_1 |A|^2 - b_1 |B|^2 ] + a_3 \bar{A} B \e^{2\i
qX} +a_4 A_{XX}, \label{eqn01} \\
\dot{B} & = & B [\mu_2 - a_2 |B|^2 - b_2 |A|^2 ] + b_3
A^2 \e^{-2\i qX} + c B_{XX}, \label{eqn02}
\ea
where the coefficients $a_j$, $b_j$ and $c$ are forced to be real
by the reflection symmetry~(\ref{reflection_sym}), the dots denote
derivatives with respect
to the slow time scale $T$ and $\mu_1$, $\mu_2$ are bifurcation
parameters. We remark that~(\ref{eqn01}) - (\ref{eqn02}) contain both
quadratic and cubic terms in the amplitudes $A$ and $B$. To ensure a
rational scheme of approximation in the limit
$\epsilon \rightarrow 0$ we should arrange that $a_3, b_3$ scale as
$\epsilon$. This can be achieved in several ways, for example where a
further symmetry which would leave the equations invariant under the
sign change $(A,B) \ra (-A,-B)$ is weakly broken. 

The form of
equations~(\ref{eqn01}) - (\ref{eqn02}) shows that when $|q|$ becomes
large the spatially-averaged contribution from the quadratic terms
decreases to zero. In this limit we formally recover the
usual `Landau' equations describing two coupled modes of instability
in the absence of spatial resonance. This is in agreement with 
our ansatz~(\ref{ansatz}); the system is far from the $1:2$ mode
interaction point when $|q| \sim O(1/\varepsilon)$. 

Subsequent calculations are made
considerably easier if we make the change of variable
$\hat{A}=A\e^{-\i qX}$ to remove the exponential factors, and rescale
the variables $A$, $B$, $T$ and $X$ to set
$a_3=a_4=1$ and $b_3=\pm 1$. Dropping the carat on $\hat{A}$ we obtain
\ba
\dot{A} & = & A [ \mu_1 - q^2 - a_1 |A|^2 - b_1 |B|^2 ] + \bar{A}B +
2\i q A_X + A_{XX}, \label{eqn1} \\
\dot{B} & = & B [ \mu_2 - a_2 |B|^2 - b_2 |A|^2 ] \pm A^2 + cB_{XX},
\label{eqn2}
\ea
which is the form of the equations that we will use in what
follows. These equations have the same structure as the ODEs derived
in Part 1 with the addition of terms giving modulation on the long
lengthscale $X$. The term $2\i q A_X$ captures the effect of
the departure from exact $1:2$ resonance. 

Writing $A=R(X,T)\e^{\i\theta(X,T)}$ and
$B=S(X,T)\e^{\i \phi(X,T)}$, the evolution equations~(\ref{eqn1}) -
(\ref{eqn2}) become
\ba
\dot{R} & = & R[\mu_1-q^2-a_1R^2-b_1 S^2] + RS \cos \newchi - 2qR\theta_X
+ R_{XX} - R(\theta_X)^2, \label{eqn_R} \\
R \dot{\theta} & = & RS \sin \newchi + 2q R_X + 2R_X \theta_X + R
\theta_{XX}, \label{eqn_theta} \\
\dot{S} & = & S[\mu_2 - a_2S^2 - b_2 R^2 ] \pm R^2 \cos \newchi + c[S_{XX}
- S(\phi_X)^2], \label{eqn_S} \\
S\dot{\phi} & = & \mp R^2 \sin \newchi + c[2S_X \phi_X + S\phi_{XX}], \label{eqn_phi}
\ea
where $\newchi\equiv \phi-2\theta$. When the 
modulational terms are omitted, we can express the dynamics in terms of only the
two moduli and the one phase difference $\newchi$ (i.e. a reduction to a
third-order system), as was done in Part 1. This leads to the ODEs
\ba
\dot{R} & = & R[\mu_1 - q^2 - a_1 R^2 - b_1 S^2] + RS \cos \newchi,
\label{eqn03} \\
\dot{S} & = & S[\mu_2 - a_2 S^2 - b_2 R^2] \pm R^2 \cos \newchi,
\label{eqn04} \\
\dot{\newchi} & = & \left( \mp \frac{R^2}{S} - 2S \right) \sin \newchi. \label{eqn05}
\ea
However, in the presence of modulational terms, each of the individual
phase variables $\theta$ and $\phi$ is dynamically
independent. Vyshkind \& Rabinovich \cite{VR76} introduced the new
variables $u=S \cos \newchi$, $v=S \sin \newchi$ (also used by Porter
\& Knobloch \cite{PK00}) to produce a third-order system avoiding the
co-ordinate singularity, but this change of variables offers no
simplification in the PDE problem.

The choice of the sign of the $\pm A^2$ term in~(\ref{eqn2})
has a huge effect on the dynamics. It was shown in Part 1 that in the
`$+$' case for the non-modulated problem the dynamics are much less
interesting than those may occur in the `$-$' case. Indeed in the
`$+$' case there is a Lyapounov functional $V(R,\theta,S,\phi)$ for
the dynamics when $2b_1=b_2$:
\ba
V  =  \left< \frac{a_1}{2}R^4 \right. & + & \frac{a_2}{4} S^4 + b_1 R^2 S^2 -
\frac{\mu_2}{2}S^2 - \mu_1 R^2 + R_X^2 + R^2(\theta_X+q)^2\\
 &-& R^2\left.
S\cos \newchi + \frac{c}{2}[S_X^2 + S^2\phi_X^2] \right>, \nn
\ea
where $\left< \cdots \right>$ denotes a horizontal average. Then a
direct calculation shows that
$$\dot{V} = - \left< 2 \dot{R}^2 + 2
R^2\dot{\theta}^2 + \dot{S}^2 + S^2 \dot{\phi}^2 \right> \leq 0,$$ i.e.
the system evolves monotonically towards a steady
state. More complicated dynamics, for example temporal oscillations,
are not possible. It follows that, at least when
$2b_1$ and $b_2$ are not widely different, or the amplitudes are
small (equivalently, near to the codimension-2 point), we expect
solution trajectories to tend asymptotically to equilibria after long
times.

For the remainder of the paper we consider the `$-$' case,
choosing the minus sign in~(\ref{eqn2}). In this case it is not
possible to construct a
Lyapounov functional, and even in the absence of modulational terms
the related ODE problem~(\ref{eqn03}) - (\ref{eqn05}) can indeed display
oscillatory dynamics; moreover there is a region of
parameter space where a robust heteroclinic cycle exists and is
stable. This cycle was analysed in Part 1 and by Armbruster et
al. \cite{AGH87}. The
parameter $c$ in~(\ref{eqn2}) corresponds to differential diffusion
rates of the two amplitudes. Varying $c$ away from unity 
leads to Turing instabilities, well-known in the context of
reaction-diffusion equations. We remove this complicating consideration by
setting $c=1$ in nearly all of what follows. The occurrence of Turing
instabilities in mode interactions is common to all cases of strong
spatial resonance and is of a different type to the instabilities we
discuss here, since it occurs in the spatially-extended but
exactly-resonance case. A full discussion of this second class of
instabilities is given in a companion paper (\cite{DP03}, in
preparation).


\section{Non-modulational dynamics near $\mu_1=\mu_2=0$}

\label{sec:analysis}

In this section we summarise the relevant parts of the bifurcation
sequences observed in Part 1 near $\mu_1=\mu_2=0$ 
in the analysis of the ODEs~(\ref{eqn03}) - (\ref{eqn05}), setting
$q=0$.
As remarked on earlier, even in the absence of the modulational terms,
a full analysis of~(\ref{eqn1}) - (\ref{eqn2}) is extremely complicated.

The trivial equilibrium $A=B=0$ is stable in the quadrant $\mu_1<0$,
$\mu_2<0$. Near the codimension-2 point at $\mu_1=\mu_2=0$, the
ODEs~(\ref{eqn03}) - (\ref{eqn05}) support simple non-trivial
equilibrium solutions of three types. The first type is a pure mode
solution $P$, of the form $A=0$, $|B|^2=\mu_2/a_2$. The
continuous symmetry of the problem implies the existence of a group
orbit of equilibria; that is, the phase of $B$ is arbitrary due to the
underlying translational symmetry. Within the subspace
$\mathrm{Fix}(m_x)$ where $A,B \in
\mathbb{R}$ there are two equilibria, denoted $P_{\pm}$. The pure mode
solutions bifurcate from the trivial solution and exist when
$\mu_2>0$. They are stable for $\mu_1$ sufficiently negative.
The other equilibria are mixed mode equilibria of two kinds,
$M_{\pm}$, corresponding to $\newchi$ taking the values  $\newchi=0$ and $\newchi=\pi$,
respectively. The $M_{\pm}$ amplitudes $R=R_0$ and $S=S_0$ are given
by solutions of the equations
\ba
0 & = & \mu_1 \pm S_0 - a_1 R_0^2 - b_1 S_0^2, \label{R0_eqn001} \\
0 & = & \mu_2 S_0 \mp R_0^2 - a_2 S_0^3 - b_2 R_0^2 S_0. \label{S0_eqn001}
\ea
where the sign choices select either $M_+$ or $M_-$. 
In fact there is a group orbit of each of $M_+$ and $M_-$ equilibria
also, since although $\newchi$ is fixed at $0$ or $\pi$ respectively,
there is a free choice of one of the underlying phases $\theta$ or
$\phi$. When $\mu_1$ is increased, holding $\mu_2>0$ fixed, stable
$M_+$ equilibria are created in a bifurcation at which the pure mode
solutions lose stability.

The $M_+$ solution then loses stability (as $\mu_1$ is increased
further) either through a symmetry-breaking drift bifurcation which produces
travelling waves (TW), or through a Hopf bifurcation to standing 
waves (SW). The TW solution resembles the mixed-mode
$M_+$ in form, but it drifts along the group orbit of mixed-mode
solutions as time evolves. The TW bifurcation is clearly a phase
instability rather than an amplitude
instability. Because the phase variables $\theta$ and $\phi$
evolve at constant rates such that $\newchi$ is constant and
$\sin \newchi \neq 0$, the TW solutions appear as equilibria in the
$(R,S,\newchi)$ variables. In other words, the use of the $(R,S,\newchi)$
variables identifies all points on the group orbit of $M_+$ solutions
as a single equilibrium, and, in these co-ordinates, information
about drift around group orbits is lost.

\begin{figure}
\begin{center}
(a) \includegraphics[width=12.0cm]{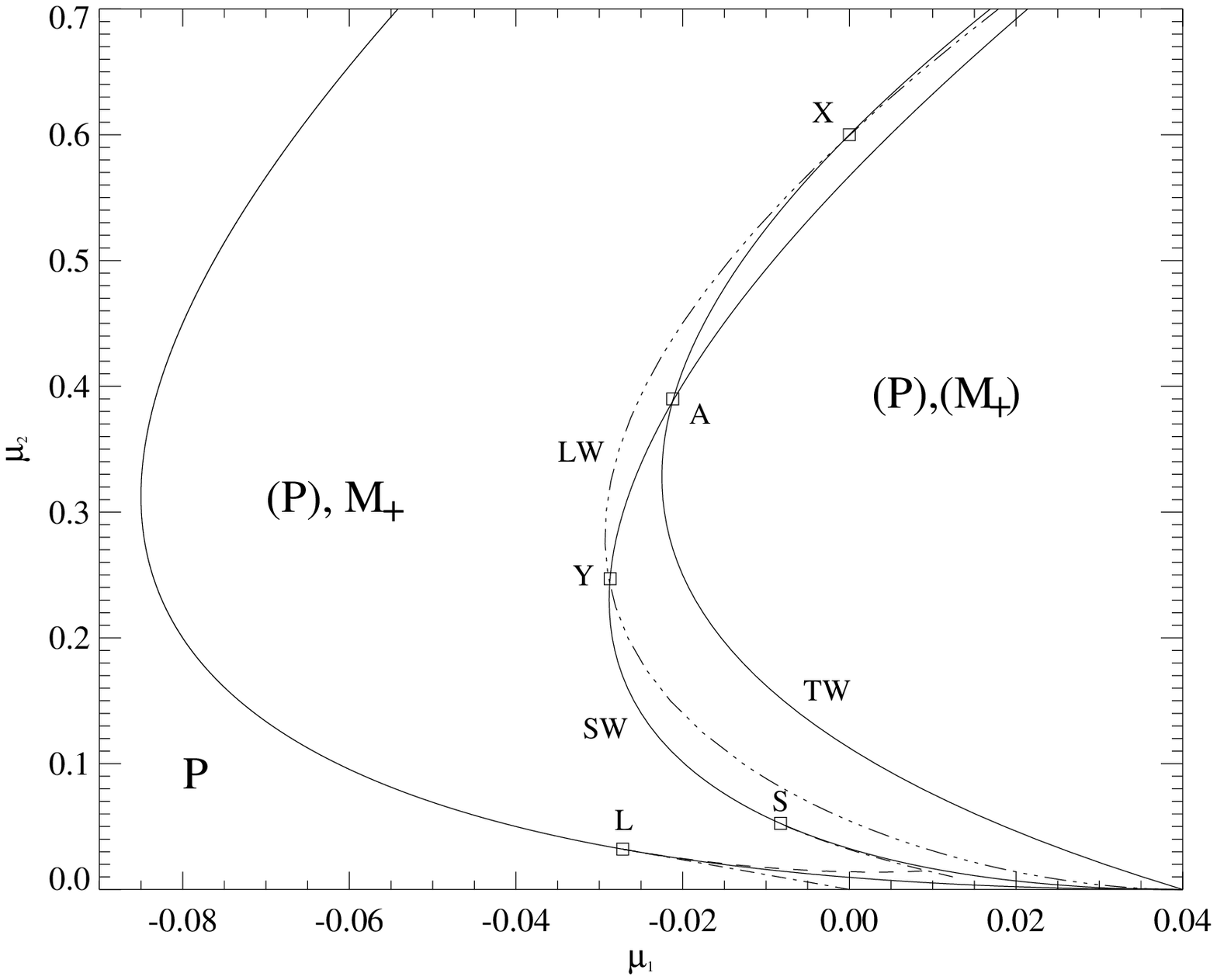}
(b) \includegraphics[width=12.0cm]{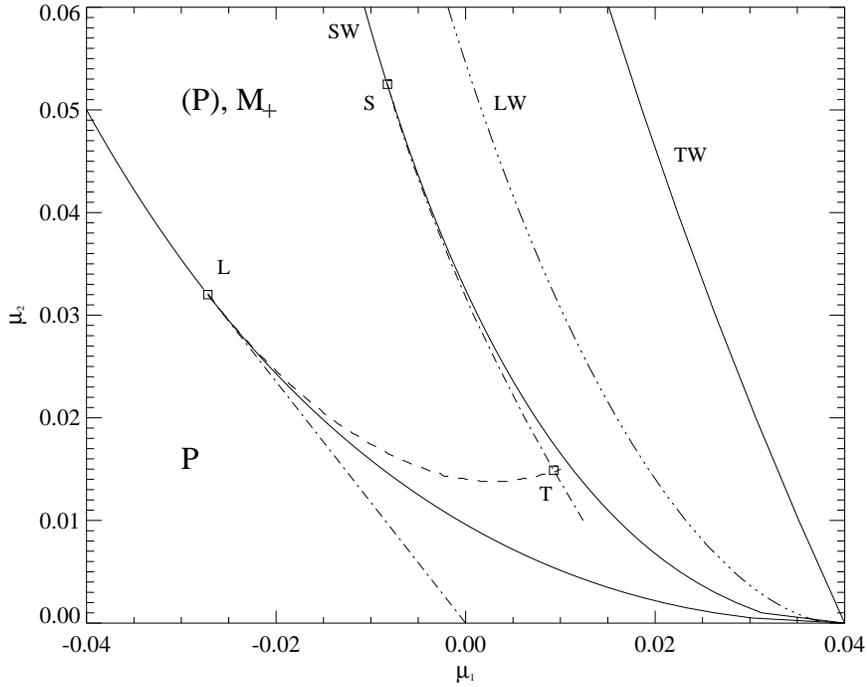}
\caption{(a) Instability boundaries in the $(\mu_1,\mu_2)$ plane for the
illustrative coefficient choices $a_1=1$, $a_2=5$, $b_1=2$, $b_2=0$,
$c=1$, $q=0.2$. Unstable and stable solutions in each region are
indicated with and without parentheses. 
The solid curve containing the point $\L$ denotes the
boundary of existence of $M_+$. The solid curves $TW$ and $SW$ denote
instabilities of $M_+$ to travelling and standing waves
respectively. The codimension-2 points $\A$, $\X$, $\Y$, $\L$, $\S$
and $\T$ and the various dashed and dash-dotted curves are
discussed in the text. (b) is an enlargement of (a).}
\label{mu1mu2_fig}
\end{center}
\end{figure}

In contrast, the SW bifurcation is an amplitude-driven instability and
produces periodic orbits that lie within the subspace $\newchi=0$. 
Typical curves along which $M_+$ undergoes bifurcations to TW or SW
solutions are shown in figure~\ref{mu1mu2_fig}. For a large region of
parameter space these two curves intersect at a codimension-2 point,
labelled $\A$ in figure~\ref{mu1mu2_fig}. Because
the eigenvectors corresponding to the TW and SW bifurcations are
orthogonal the codimension-two bifurcation corresponding to
simultaneous instability is a pitchfork--Hopf bifurcation (in the
$(R,S,\newchi)$ co-ordinates).

As we increase $\mu_1$ for small positive $\mu_2$, the SW
instability occurs first. Within the subspace $\mathrm{Fix}(m_x)$ the
periodic orbit created in the SW
bifurcation grows until it collides simultaneously with the origin and
the $P_+$ equilibrium. After this global bifurcation the periodic orbit
disappears, but now the unstable manifold of $P_+$ tends
asymptotically to the pure mode solution $P_-$. By symmetry, within
the invariant subspace $\mathrm{Fix}(m_x \circ \tau_\pi)$ the
unstable manifold of $P_-$ tends to $P_+$, and a heteroclinic cycle is
formed. The heteroclinic
cycle is structurally stable due to the existence of
invariant subspaces, within which each connecting trajectory
lies. Hence it exists for an open set of values of the coefficients
and bifurcation parameters. Moreover, this cycle is attracting for an
open interval of values of $\mu_1$.

At larger
$\mu_1$ this cycle ceases to attract nearby trajectories as it
undergoes a resonant bifurcation, resulting in a global loss of
stability, and creating modulated waves (MW). MW
are destroyed in a Hopf bifurcation from the TW which themselves cease
to exist in a bifurcation with the $M_-$ states. Moreover, 
several other heteroclinic cycles are possible for other combinations of
coefficients, and at larger positive $\mu_2$. These have been
investigated in detail by Porter \& Knobloch \cite{PK00}.


\section{Modulational instability of the pure mode}

\label{sec:pure_mode}

Having summarised the behaviour of the system in the absence of
spatial modulations, we now allow $q$ to take non-zero values, and
examine the possibility of new modes of instability due to the spatial
frequency mismatch. We first note that the results of Part 1,
summarised in the previous section, apply 
within the subspace of non-modulated ($X$-independent) solutions.
To examine the stability of the pure mode solution $P_+$, given by
$A=0$ and $B=B_0\equiv\sqrt{\mu_2/a_2}$, we substitute the ansatz
\ba
A(X,T) & = & \alpha_1(T) \e^{\i\ell X}    + \bar{\alpha}_2(T) \e^{-\i\ell
X}, \nn \\
B(X,T) & = & B_0(1+\beta_1(T) \e^{\i m X} + \bar{\beta}_2(T) \e^{-\i m X}),
\nn
\ea
into~(\ref{eqn1}) - (\ref{eqn2}) and
linearise. The linearised dynamics for $\alpha_{1,2}$ and $\beta_{1,2}$
decouple and $P_+$ is found to be stable to the perturbations
$\beta_{1,2}$ for all wavenumbers $m$. The linearised system for the
perturbations $\alpha_{1,2}$ is
\ba
\left( 
 \begin{array}{c}
  \dot{\alpha}_1 \\ 
  \dot{\alpha}_2 
 \end{array} 
\right)
& = & 
\left( 
 \begin{array}{cc} 
  \mu_1 - q^2 - b_1 B_0^2 - 2q\ell - \ell^2 & B_0 \\
  B_0 & \mu_1 - q^2 - b_1 B_0^2 + 2q\ell - \ell^2
 \end{array} 
\right)
\left( 
 \begin{array}{c}
  \alpha_1 \\ 
  \alpha_2 
\end{array} \right). \nn \\ & & \label{pplus_linear}
\ea
This matrix has trace $\mathrm{tr}(\ell^2)$ and determinant $D(\ell^2)$:
\ba
\mathrm{tr}(\ell^2) & = & 2(\mu_1 - q^2 - \ell^2 - b_1 B_0^2), \nn \\
D(\ell^2) & = & (\mu_1 - q^2 - \ell^2 - b_1 B_0^2)^2 - 4q^2 \ell^2 -
B_0^2. \nn
\ea
No oscillatory bifurcation is possible since $\mathrm{tr}(\ell^2)=0$
implies $D(\ell^2)<0$. A steady-state instability to perturbations with
wavenumber $\ell$ occurs when $D(\ell^2)=D'(\ell^2)=0$ since
$D''(\ell^2)=2>0$. These conditions show that the first instability of
$P_+$ may be to perturbations either with $\ell=0$ or with $\ell$
non-zero. When
$\mu_2>a_2(\mu_1+q^2)/b_1$ the first instability is to $\ell=0$ and occurs
along the curve $\mu_1=b_1\mu_2/a_2 + q^2 - \sqrt{\mu_2/a_2}$. When
$\mu_2<a_2(\mu_1+q^2)/b_1$ the first instability is to finite $\ell=\ell_c>0$ and
occurs along the line 
\ba
\mu_1 & = & \frac{\mu_2}{a_2} \left( b_1 - \frac{1}{4q^2} \right). \nn
\ea
These two instability curves meet at the point
$(\mu_1^\L,\mu_2^\L)=(4b_1q^4-q^2,4a_2q^4)$, marked $\L$ on
figure~\ref{mu1mu2_fig}. The gradients of these curves are equal here,
so the transition between instabilities is smooth. Along the line
$\Origin\L$ (from the origin to $\L$) the most unstable wavenumber,
$\ell_c$, is given by
$\ell_c^2=\mu_1-b_1\mu_2/a_2 + q^2$, i.e. $0 < \ell_c < q$,
and $\ell_c$ increases monotonically as the origin is approached.

At any point on the interior of the line $\Origin\L$ we can fully
describe this bifurcation by the usual
Ginzburg--Landau equation, since for the PDEs~(\ref{eqn1}) -
(\ref{eqn2}) this is an instability of a uniform state to a
nonzero-wavelength perturbation, and the
growth rate of a perturbation with a wavenumber far from $\ell_c$ is
negative and bounded away from zero. We have carried out a weakly
nonlinear perturbation expansion near $\Origin\L$ to investigate whether
this bifurcation is subcritical or supercritical, details of which
are given in Appendix 1. The resulting
analytic expression is cumbersome, but it is possible to deduce that the
bifurcation is always supercritical when $a_1>0$ is large
enough. This calculation can be carried out analytically without
assuming $c=1$; if $c$ is large compared to unity
and $a_1$ is small it is possible for the bifurcation to be
subcritical. For the illustrative set of coefficients used in
figure~\ref{mu1mu2_fig} the bifurcation is supercritical along the
whole of $\Origin\L$.

The dynamics in a neighbourhood of the
codimension-2 point $\L$ cannot, though, be described by the
Ginzburg--Landau equation, since the instability wavelength $\ell_c$
tends to zero as $\L$ is approached. A codimension-2 point identical in
structure to $\L$ occurs in the analysis by Coullet \& Repaux
\cite{CR87} of a pattern-forming
instability subjected to an external nearly-resonant periodic
forcing. They term $\L$ a `Lifschitz
point'. Through asymptotic expansions near $\L$ it is possible to
describe the dynamics in terms of a single `normal form' equation
\ba
\frac{dA}{d\tau} & = & \nu_1 A - A^3 + \nu_2 A_{\xi\xi} - A_{\xi\xi\xi\xi},
\label{EFK_eqn}
\ea
which governs the evolution of a small perturbation $A(\xi,\tau) \in
\mathbb{R}$ to the $P_+$ solution; $\nu_1$ and $\nu_2$ are new
bifurcation parameters, defined so that the point $\L$ corresponds to
$\nu_1=\nu_2=0$,
$\tau$ is a new scaled time variable and $\xi$ is a new long
spatial scale associated with the smallness of $\ell_c$. 
Equation~(\ref{EFK_eqn}) is known as the Extended
Fisher--Kolmogorov equation and it's properties have been extensively
investigated
(\cite{CER87,PT96,B00}). It is easily shown that there is a Lyapounov
functional for the dynamics, and so there are only steady state
solutions at long times. However, these states need not be periodic in
space; and in fact the solutions can have very complex spatial
structure when $\nu_1>0$, corresponding to the region
$\mu_1 > b_1 B_0^2 - B_0 + q^2$.


\section{Modulational instabilities of the mixed modes}

\label{sec:mixed_mode}

$M_+$ undergoes a plethora of different bifurcations. In this section
we will discuss three new instabilities to long-wavelength
disturbances that occur. We examine a
codimension-two bifurcation where two of these curves meet. We also
discuss other codimension-two points that occur where
one of these long-wavelength instabilities meets a bifurcation curve
from the non-modulated problem. Our discussion is organised by 
the sequence in which these bifurcations appear in
figure~\ref{mu1mu2_fig} as $\mu_1$ increases.

Let $A=R_0(1 + \alpha_1 \e^{\i\ell X} + \bar{\alpha}_2 \e^{-\i\ell X})$
and $B=S_0(1 +  \beta_1 \e^{\i\ell X} +  \bar{\beta}_2 \e^{-\i\ell
X})$ where $R_0$ and $S_0$ satisfy
\ba
0 & = & \mu_1 - q^2 + S_0 - a_1 R_0^2 - b_1 S_0^2, \label{mm_eqn1} \\
0 & = & \mu_2 S - R_0^2 - a_2 S_0^3 - b_2 R_0^2 S_0. \label{mm_eqn2}
\ea
After substituting into~(\ref{eqn1}) - (\ref{eqn2}), linearising and
changing to the sum and difference variables $\alpha_\pm = \alpha_1\pm
\alpha_2$, $\beta_\pm = \beta_1 \pm
\beta_2$, we obtain the linearisation matrix
\ba
\left(
\begin{array}{c}
R_0\dot{\alpha}_+ \\
S_0\dot{\beta}_+ \\
R_0\dot{\alpha}_- \\
S_0\dot{\beta}_-
\end{array}
\right)
& = & 
\left(
\begin{array}{cccc}
-2a_1 R_0^2 - \ell^2 & R_0 (1- 2 b_1 S_0) & -2q\ell & 0 \\
-2R_0(1+ b_2 S_0) & \frac{R_0^2}{S_0} - 2a_2 S_0^2 - c\ell^2 & 0 &
0 \\
-2q\ell & 0 & -2S_0 - \ell^2 & R_0 \\
0 & 0 & -2R_0 & \frac{R_0^2}{S_0} - c\ell^2
\end{array}
\right)
\left(
\begin{array}{c}
R_0 \alpha_+ \\ S_0 \beta_+ \\ R_0\alpha_- \\ S_0\beta_-
\end{array}
\right). \nn \\ & & \label{mplus_stab}
\ea
The characteristic polynomial of this matrix can be written as
$P(\lambda) = \lambda^4 + \hat{A}(\ell^2)\lambda^3 +
\hat{B}(\ell^2)\lambda^2 + \hat{C}(\ell^2)\lambda
+ \hat{D}(\ell^2)$. Note that $P(\lambda)$ always has a root $\lambda=0$ when
$\ell=0$, hence we may write $\hat{D}(\ell^2)=\ell^2 \hat{E}(\ell^2)$ where
$\hat{E}(\ell^2)$ is a cubic polynomial. This is due to the
underlying translation symmetry. Steady-state instabilities at
$\ell=\ell_*>0$ occur when
$\hat{E}(\ell_*^2)=\hat{E}'(\ell_*^2)=0$ and
$\hat{E}''(\ell_*^2)>0$. Oscillatory instabilities
occur when $\lambda\equiv\alpha(\ell^2)+\i\omega(\ell^2)$ satisfies
$\alpha(\ell_*^2)=\alpha'(\ell_*^2)=0$, $\alpha''(\ell_*^2)>0$ and
$\omega^2(\ell_*^2)>0$, where $\alpha(\ell^2)$ and $\omega(\ell^2)$
are real. These requirements yield the conditions 
\ba
\hat{C}^2 -\hat{A}\hat{B}\hat{C} + \hat{A}^2 \hat{D} & = & 0, \nn \\
(\hat{A} \hat{C}' - \hat{C}\hat{A}')(2\hat{C} -
\hat{A}\hat{B}) - \hat{A}^2(\hat{B}'\hat{C} - \hat{A}\hat{D}') & = & 0, \nn \\
\hat{C}/\hat{A} & > & 0, \nn
\ea
for an oscillatory bifurcation with $\ell_*>0$.
Both the steady-state and the oscillatory cases give two conditions; 
in conjunction with~(\ref{mm_eqn1}) - (\ref{mm_eqn2}), these
conditions enable the determination of bifurcation lines in the
$(\mu_1,\mu_2)$ plane. Clearly, a
steady-state instability at $\ell_*=0$ occurs when $\hat{E}(0)=0$ and
$\hat{E}'(0)>0$; similarly an oscillatory instability occurs when
$\alpha(0)=0$ and $\alpha'(0)<0$ as long as $\omega^2(0)>0$. In terms
of the coefficients of $P(\lambda)$ these conditions are
\ba
\hat{C}^2 - \hat{A}\hat{B}\hat{C} + \hat{A}^2 \hat{D} & = & 0, \nn \\
\frac{(\hat{B}\hat{C}-\hat{A}\hat{D}-\hat{C}^2/\hat{A})'}{2(4\hat{D}-\hat{A}\hat{C}-\hat{B}^2)}
& < & 0, \nn \\
\hat{C}/\hat{A} & > & 0. \nn
\ea

\subsection{Instability of $M_+$ near the Lifschitz point $\L$}

\label{mplus_l}

The first new instability of $M_+$ that we find is a steady-state
bifurcation to spatially-modulated solutions, i.e. the most unstable
wavenumber is non-zero. This occurs
along the dashed curve $\L\T$ emanating from the Lifschitz point $\L$. This
instability is part of the generic bifurcation structure near a Lifschitz
point and hence the existence of this curve can be shown by analysis
of the normal form~(\ref{EFK_eqn}). We have followed the
bifurcation curve to larger values of $\mu_1$ where it becomes
asymptotic to the line $\mu_1=q^2$ as $\mu_2 \ra \infty$. 
Near $\L$, when $a_1$ is large we expect the bifurcation
to be subcritical by comparison with the results of Coullet \& Repaux
\cite{CR87}. This bifurcation curve meets a second such
non-zero-$\ell$ instability curve (this one starting from the point
$\S$) at the point $\T$ on figure~\ref{mu1mu2_fig}. 
\begin{figure}
\begin{center}
\includegraphics[width=10.0cm]{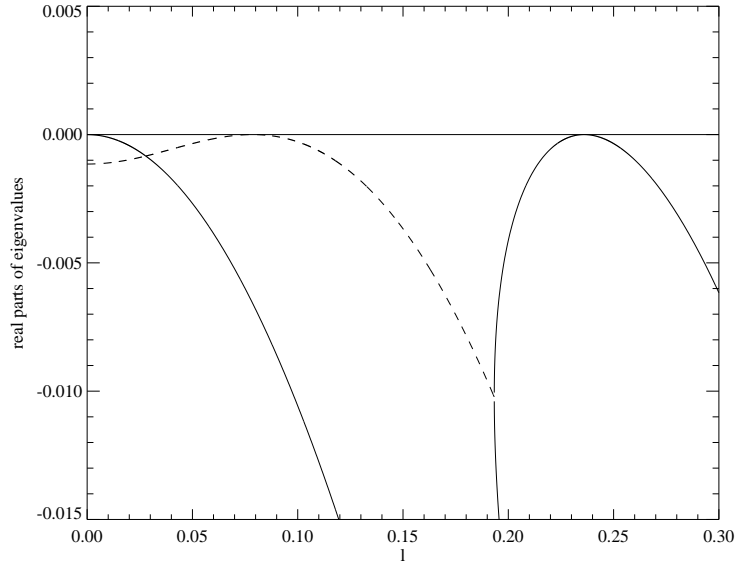}
\caption{Codimension-two instability of $M_+$ at the point $\T$,
$(\mu_1,\mu_2)=(0.00929,0.0149)$ for $q=0.2$ and the coefficients of
figure~\ref{mu1mu2_fig}. Real 
parts of the eigenvalues $\lambda$ of the linearisation about $M_+$
are plotted against wavenumber $\ell^2$. Solid lines denote real
eigenvalues, dashed lines give the real part of complex conjugate
pairs. Note the mode with zero growth rate at $\ell=0$ due to the
translational symmetry of $M_+$.}
\label{point_T_fig}
\end{center}
\end{figure}
At $\T$ there is a Hopf/steady-state mode
interaction coupled to a phase mode which has zero
growth rate at zero wavenumber; figure~\ref{point_T_fig} shows the
variation of the real parts of the eigenvalues $\lambda(\ell^2)$ at
this point. The dynamics near the codimension-two point $\T$ depend
strongly on the ratio of the wavenumbers involved in the
instabilities.
\begin{figure}
\begin{center}
\includegraphics[width=6.7cm]{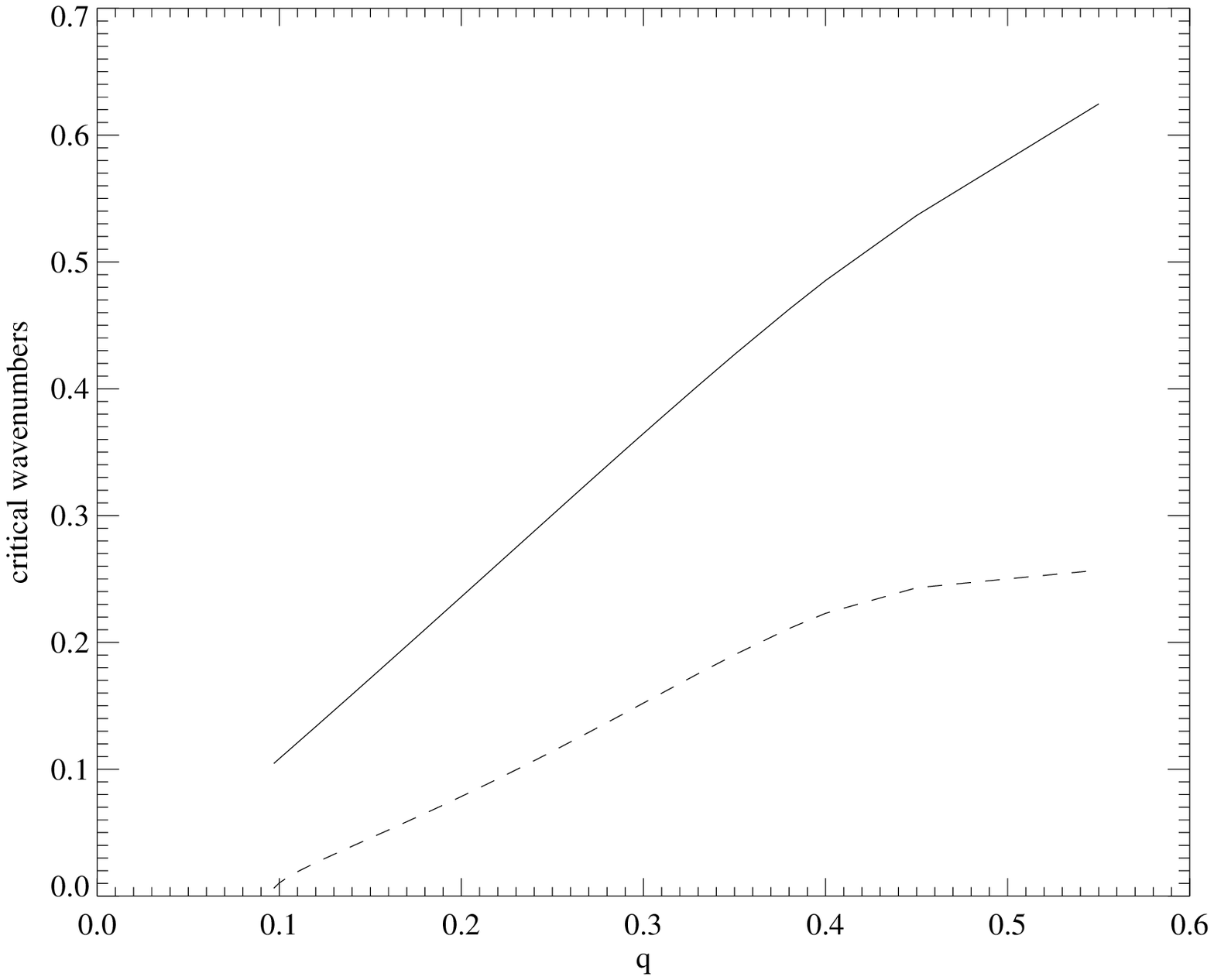}
\includegraphics[width=6.7cm]{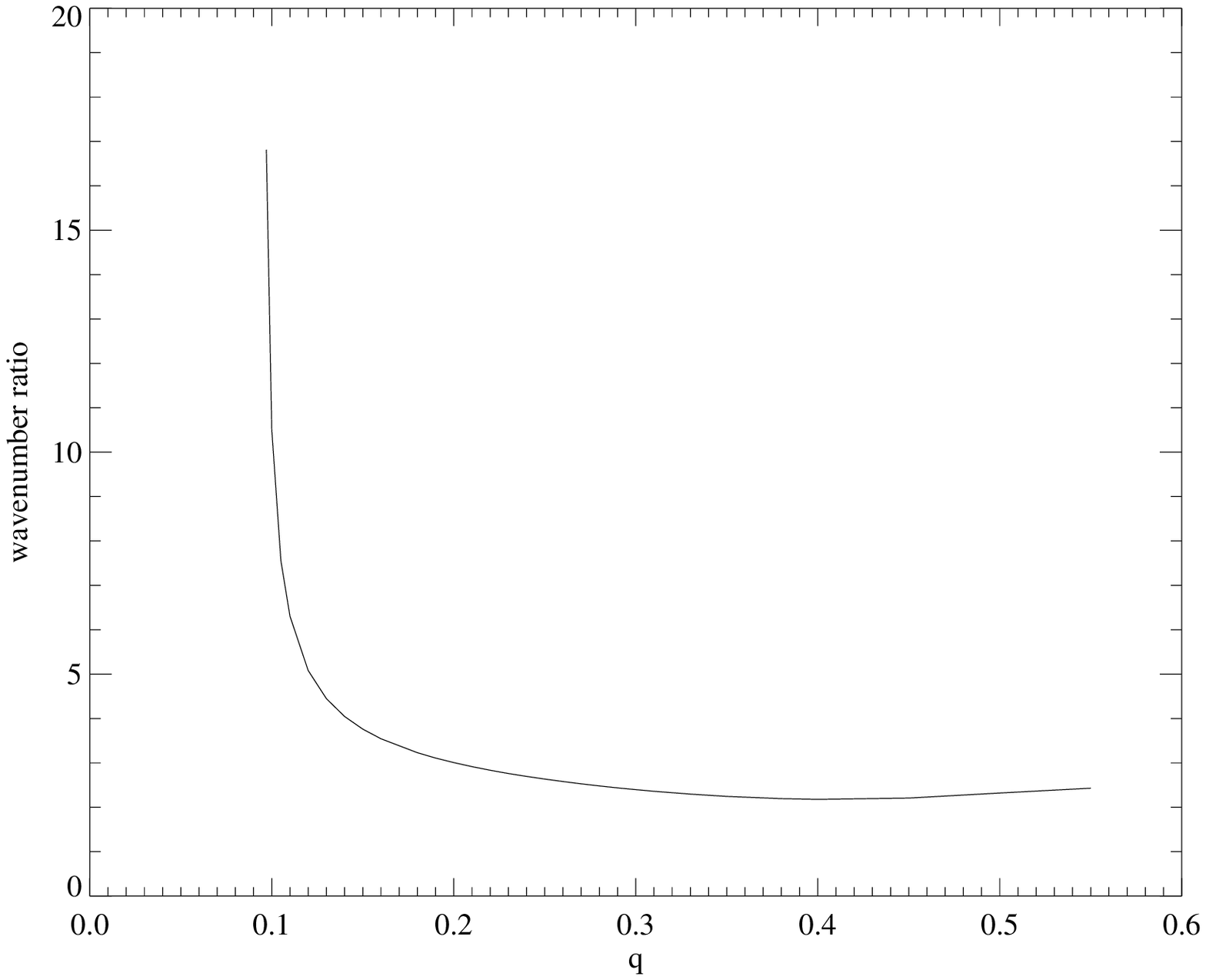}
\caption{(a) Critical wavenumbers for the steady-state (solid line)
and oscillatory (dashed line) instabilities at $\T$ (see
figure~\ref{mu1mu2_fig}) as a function of the
wavenumber mismatch $q$, in the range $0.097<q<0.55$. (b) Ratio of the
critical wavenumbers in (a) as a function of $q$.}
\label{point_T_ratio_fig}
\end{center}
\end{figure}
This ratio varies considerably with $q$, as shown in
figure~\ref{point_T_ratio_fig}. For the illustrative coefficient set
used in figure~\ref{mu1mu2_fig}, when $q=0.097$ there is a
codimension-three bifurcation as $\S$ collides with $\T$. As we
approach this collision, the wavenumber of the oscillatory instability
at $\T$ goes to zero and so the wavenumber ratio is formally infinite
there.


The critical wavenumbers for the
steady-state instability can be seen from
figure~\ref{point_T_ratio_fig}(a) to be slightly 
greater than the mismatch parameter $q$, in contrast with the
modulational instability of the pure mode 
solution $P_+$ discussed in section~\ref{sec:pure_mode} where the
maximum critical wavenumber of instability is exactly $q$. It turns
out that for $M_+$ the maximum instability wavenumber is $2q$. This
can be easily shown by examining $M_+$ near the line $\mu_1=q^2$,
$\mu_2>0$. When $\mu_1-q^2$ is small and negative the amplitudes $R_0$
and $S_0$ for $M_+$ are, solving~(\ref{mm_eqn1}) - (\ref{mm_eqn2}) to
leading order:
\ba
R_0^2 & = & \frac{(\mu_1-q^2)\mu_2}{a_1 \mu_2 -1}, \nn \\
S_0 & = & \frac{\mu_1-q^2}{a_1 \mu_2 -1}. \nn
\ea
Then, substituting these expressions into the term $\hat{E}(\ell^2)$
from the characteristic polynomial $P(\lambda)$ we obtain
\ba
\hat{E}(\ell^2) & = & (\mu_2 - \ell^2)^2(\ell^2-4q^2) + O(|\mu_1-q^2|). \nn
\ea
The double root at $\ell^2=\mu_2$ comes from the LW instability; it is
the root at $\ell^2=4q^2$ that comes from the continuation of the
steady-state instability curve through $\T$ towards $\mu_1-q^2 \approx
0$. From numerical investigations we conjecture that the wavenumber of
the instability evolves monotonically along the curve, but there does
not seem to be a straightforward way to verify this analytically.

\subsection{Instability of $M_+$ to standing waves}

The dash-dotted line $\T\S$ in figure~\ref{mu1mu2_fig}(b) marks the
instability boundary of $M_+$ to the second new bifurcation involving
spatial modulation; an oscillatory instability to temporal 
oscillations (standing waves) with non-zero spatial wavenumber.
The dynamics near the point $\S$ is
analogous (but time-periodic rather than steady) to that near the
Lifschitz point $\L$, and can thus be described by a similar extension
of the complex Ginzburg--Landau equation:
\ba
\frac{dA}{d\tau} & = & (\nu_1+\i\omega_0(\nu_1,\nu_2)) A - (1+\i\beta)A|A|^2 +
(\nu_2+\i\alpha)  A_{\xi\xi} - (1+\i\gamma) A_{\xi\xi\xi\xi}, \nn \\ &
& \label{EFK_eqn002}
\ea
governing the evolution of a complex-valued perturbation
$A(\xi,\tau)$ to the $M_+$ solution. As
before, $\nu_1$ and $\nu_2$ are new
bifurcation parameters; the point $\S$ corresponds to
$\nu_1=\nu_2=0$, $\omega_0$ is the non-zero frequency of the
instability, $\alpha$, $\beta$ and $\gamma$ are real parameters,
$\tau$ is a new scaled time variable and $\xi$ is a new long
spatial scale. Clearly the dynamics of~(\ref{EFK_eqn002}) are at least
as complicated as those of~(\ref{EFK_eqn}). The wavenumber of this
oscillatory instability increases as $\mu_2$ decreases along the curve
(from $\S$ to $\T$). 



\subsection{The long wavelength phase instability of $M_+$}

The third new instability occurs, at larger $\mu_2$, along the curve LW
in figure~\ref{mu1mu2_fig}(a). This instability is a long-wavelength
steady-state bifurcation; as $q$ is increased from zero, the LW curve
splits off from the TW curve along its entire length. Since, in the
absence of modulational terms, $M_+$ are unstable first to SW below
$\A$ and unstable first to TW above $\A$, the LW instability
(at least for this combination of coefficients) is the instability of
$M_+$ that occurs first in an intermediate range of $\mu_2$, between
the new codimension-two points $\X$ and $\Y$. The LW and SW curves
intersect at $\Y$ and the LW and TW curves intersect at $\X$, see
figure~\ref{mu1mu2_fig}(a).
Since the LW instability curve coalesces with the TW instability as $q
\ra 0$ it is clear that the LW instability must also be a phase
instability rather than an amplitude instability.

It is important to observe that the point $\A$, where the TW and SW
curves cross, is now `shielded' by the LW curve. The dynamics near $\A$
were investigated in detail in this context by Julien \cite{J91}, and
in greater generality by Landsberg \&
Knobloch \cite{LK93}. In the presence of
modulations the dynamics near $\A$ are expected to be less relevant to
the observed dynamics in a spatially-extended physical system.
However, it is worth noting that by varying the diffusivity
ratio $c$ away from unity it is possible to make the points $\X$ and
$\A$ coincide; this would result in a codimension-three bifurcation
involving the TW, LW and SW instabilities of $M_+$.

Reduced descriptions of the dynamics near $\X$ and $\Y$ can be derived 
from the PDEs starting either from~(\ref{eqn1}) - (\ref{eqn2}) or,
more conveniently here, from
the modulus/phase representation of the dynamics given
by~(\ref{eqn_R}) - (\ref{eqn_phi}). In this subsection we will present
the reduction near $\X$ in some detail, and comment only briefly on the
dynamics near $\Y$.

Near $\X$ we strive to eliminate the equations~(\ref{eqn_R})
and~(\ref{eqn_S}) for the evolution of the moduli $R$ and $S$ by
adiabatic
elimination to leave a pair of real equations for the phases $\theta$
and $\phi$. It turns out to be more convenient to describe the
dynamics in terms of $\newchi\equiv\phi-2\theta$ and $\theta$, since the
TW instability involves only the combination
$\newchi$. From~(\ref{eqn05}) it is clear that the TW instability occurs
when
\ba
R_0^2 & = & 2S_0^2, \label{TW_condition}
\ea
where $R_0$ and $S_0$ satisfy~(\ref{mm_eqn1}) - (\ref{mm_eqn2}). The
LW instability occurs when
$\hat{E}(0)=0$ where $\ell^2\hat{E}(\ell^2)$ is the determinant of the
linearisation matrix~(\ref{mplus_stab}):
\ba
\hat{E}(0) & = & \frac{2R_0^2}{S_0}\left[ (2cS_0^2-R_0^2)(2(a_1
a_2-b_1 b_2)S_0^3 + (b_2-2b_1)S_0^2 \right. \nn \\ 
& & \left. - a_1R_0^2 + S_0) + 2q^2(2a_2S_0^3-R_0^2) \right]. \label{LW_condition}
\ea
Note that when $c=1$ and $q=0$ this condition reduces to the condition for the
TW instability~(\ref{TW_condition}). When $c=1$ the algebra simplifies
substantially (and so we set $c=1$ for the remainder of this section);
solving~(\ref{TW_condition}) and (\ref{LW_condition}) together
yields the moduli $R_0=\sqrt{2}/a_2$, 
$S_0=1/a_2$ at the point $\X$ where the LW and TW instabilities
coincide. Substitution into~(\ref{mm_eqn1}) - (\ref{mm_eqn2}) gives
the co-ordinates of the point $\X$ in the $(\mu_1,\mu_2)$ plane:
\ba
(\mu_1^\X,\mu_2^\X) & = & ((2a_1+b_1-a_2)/a_2^2+q^2,(3a_2+2b_2)/a_2^2). \nn
\ea
The adiabatic elimination of $R$ and $S$ at the point $\X$ proceeds in
the usual manner, but the scaling for $\newchi$ differs from that which
might be expected. Our choices of scalings are determined completely
by the requirement to balance the linear terms in the
reduced equations. This balance then introduces two specific nonlinear
coupling terms at leading order. We write
\ba
R & = R_0 + \epsilon^2 r_0 + \epsilon^4 r_1 + \cdots, \ \ \
\tilde{X} & = \epsilon X, \label{scale_1} \\
S & = S_0 + \epsilon^2 s_0 + \epsilon^4 s_1 + \cdots, \ \ \ 
\tilde{T} & = \epsilon^2 T, \label{scale_2} \\
(\theta,\phi) & = \epsilon (\tilde{\theta},\tilde{\phi}), \hspace{3.2cm}
\newchi & = \epsilon^3 \tilde{\newchi}, \label{scale_3} \\
\mu_1 & = \mu_1^\X + \epsilon^2 \hat{\mu}_1, \hspace{2.5cm}
\mu_2 & = \mu_2^\X + \epsilon^2 \hat{\mu}_2. \label{scale_4}
\ea
Substituting into~(\ref{eqn_R}) and~(\ref{eqn_S}) and dropping
the tildes gives, at $O(\epsilon^2)$:
\ba
2a_1 R_0^2 r_0 + (2b_1 S_0 - 1)R_0 s_0 & = & R_0 \hat{\mu}_1 - 2 q R_0 \theta_X, \nn \\
2 R_0(1 + b_2 S_0) r_0 & = & S_0 \hat{\mu}_2. \nn
\ea
Hence
\ba
r_0 & = & \frac{S_0 \hat{\mu}_2}{2R_0(1+b_2 S_0)}, \label{r0_eqn} \\
s_0 & = & \frac{\hat{\mu}_1-2q\theta_X}{2b_1S_0-1} - \frac{S_0
\hat{\mu}_2}{(2b_1 S_0 - 1)(1 + b_2 S_0)}. \label{s0_eqn}
\ea
Substituting the scalings~(\ref{scale_1}) - (\ref{scale_4}) into the
$\dot{\theta}$ equation~(\ref{eqn_theta}), dropping the tildes and
cancelling a factor of $\epsilon^3$ yields
\ba
R_0 \dot{\theta} & = & R_0 S_0 \newchi + 2 q r_{0X} + R_0
\theta_{XX} + O(\epsilon^2). \label{theta_eqn_lw001}
\ea
From~(\ref{r0_eqn}) we see that $r_{0X}=0$ and so~(\ref{theta_eqn_lw001})
simplifies to give
\ba
\dot{\theta} & = & S_0 \newchi + \theta_{XX} + O(\epsilon^2), \label{theta_eqn_lw}
\ea
which is the first of our pair of reduced equations.
It turns out that we need to compute the term $r_1$ to determine the
leading order evolution of $\newchi$ correctly. From~(\ref{eqn_S})
at $O(\epsilon^4)$ we find
\ba
2 R_0(1 + b_2 S_0) r_1 & = & s_0 \hat{\mu}_2 - 3a_2S_0 s_0^2 - (1+b_2
S_0) r_0^2 - 2b_2 R_0 r_0 s_0 \nn \\ 
& & + s_{0XX} - S_0 (\phi_X)^2 - \dot{s}_0,
\nn
\ea
which, after substituting~(\ref{r0_eqn}) and~(\ref{s0_eqn}), becomes
\ba
2 R_0(1 + b_2 S_0) r_1 & = & \left( \frac{12q a_2 S_0
[(1+b_2S_0)\hat{\mu}_1-S_0\hat{\mu_2}]}{(2b_1 S_0-1)^2(1+b_2 S_0)}
- \frac{2 q \hat{\mu}_2}{(1 + b_2 S_0)(2b_1 S_0 - 1)}
\right) \theta_X \nn \\ 
& & - 4 S_0 (\theta_X)^2 + \frac{2q}{2b_1S_0-1}(\dot{\theta}_X -
\theta_{XXX}) + \mathit{const}, \label{r1_eqn}
\ea
where $\mathit{const}$ denotes terms independent of $X$, and the last
term is equal to $2qS_0 \newchi_X/(2 b_1 S_0-1)$ at leading
order, using~(\ref{theta_eqn_lw}).

\begin{figure}
\begin{center}
\includegraphics[width=7.0cm]{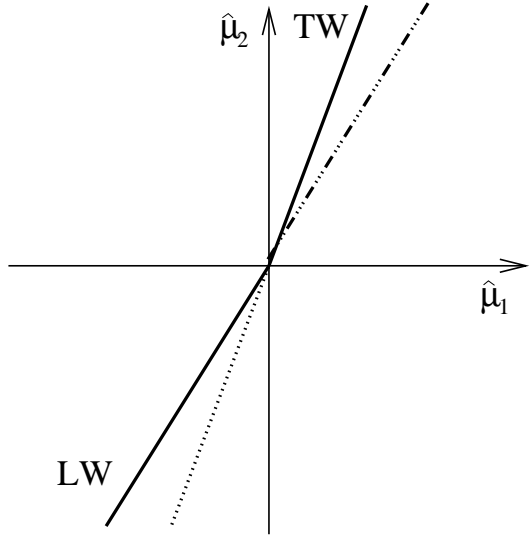}
\caption{Sketch bifurcation lines near the codimension-two point $\X$
where the TW and LW curves meet, for the illustrative coefficient
choices $a_1=1$, $a_2=5$, $b_1=2$, $b_2=0$,
$c=1$, $q=0.2$. $M_+$ are stable above and to the left of the LW and
TW curves. The TW bifurcation (solid/dotted line) occurs along the
line $20\hat{\mu}_1-3\hat{\mu}_2=0$; it is the initial instability of
$M_+$ when $6\hat{\mu}_1-\hat{\mu}_2>0$
(indicated by the solid line). The dotted part of the line indicates
instability in the absence of long-wavelength modulations. The LW
bifurcation (solid/dash-dot-dot-dotted line) occurs similarly when
$(20+1500q^2)\hat{\mu}_1=(3+250q^2)\hat{\mu}_2$ and
$20\hat{\mu}_1<3\hat{\mu}_2$. The solid part of the LW line indicates
that part of it for which it is the initial instability of $M_+$.}
\label{point_X_fig}
\end{center}
\end{figure}

Now we turn to the (unscaled) $\dot{\newchi}$ equation, formed by
combining~(\ref{eqn_theta}) and~(\ref{eqn_phi}):
\ba
\dot{\newchi} & = & \left( \frac{R^2}{S} - 2S \right) \newchi + \newchi_{XX} +
\frac{2}{S}S_X \phi_X - \frac{4q}{R}R_X - \frac{4}{R}R_X \theta_X. \nn
\ea
After substituting the scalings~(\ref{scale_1}) - (\ref{scale_4}) and
dropping the tildes, the terms at $O(\epsilon^5)$ are found to be
\ba
\dot{\newchi} & = & \left[ \frac{2R_0}{S_0}r_0 - \left(
2+\frac{R_0^2}{S_0^2} \right) s_0 \right] \newchi 
+ \newchi_{XX} + \frac{4}{S_0}s_{0X}\theta_X - \frac{4q}{R_0}r_{1X},
\label{chi_eqn_lw001}
\ea
using the fact that $r_{0X}=0$. After substituting for $s_{0X}$ and
$r_{1X}$ using~(\ref{s0_eqn}) and~(\ref{r1_eqn}) (noting that the
terms in~(\ref{r1_eqn}) indicated by `\emph{const}' do not appear) we
obtain
\ba
\dot{\newchi} & = & \xi_0 \newchi + (1+\xi_4) \newchi_{XX} + \xi_1 \theta_{XX}
+ \xi_2 \newchi \theta_X + \xi_3 \theta_X \theta_{XX} + O(\epsilon^2), \label{chi_eqn_lw}
\ea
where
\ba
\xi_0 & = & \hat{\mu}_2 \left[\frac{a_2}{a_2+b_2}+
\frac{4a_2}{(2b_1-a_2)(a_2+b_2)} \right] -\frac{4a_2
\hat{\mu}_1}{2b_1-a_2}, \nn \\
\xi_1 & = & \frac{-12q^2 a_2^5 \hat{\mu}_1}{(a_2+b_2)(2b_1-a_2)^2} 
+ \frac{2 q^2 a_2^5 \hat{\mu}_2(6+2b_1-a_2)}{(a_2+b_2)^2(2b_1-a_2)^2}, \nn \\
\xi_2 & = & \frac{8 q a_2}{2b_1 - a_2}, \nn \\
\xi_3 & = & \frac{4q a_2^2(2b_1-3a_2-2b_2)}{(a_2+b_2)(2b_1-a_2)}, \nn \\
\xi_4 & = & \frac{-2q^2 a_2^3}{(a_2+b_2)(2b_1-a_2)}. \nn
\ea
One simple consistency check is that coefficients containing odd
powers of $q$ multiply terms with odd numbers of $X$-derivatives (and
likewise for even powers of $q$).
For the illustrative coefficient choices $a_1=1$, $a_2=5$, $b_1=2$ and $b_2=0$
we obtain the leading order reduced equations
\ba
\dot{\newchi} & = & (20\hat{\mu}_1-3\hat{\mu}_2) \newchi + (1+50q^2) \newchi_{XX} 
\nn \\ 
& & - 1250q^2(6\hat{\mu}_1-\hat{\mu}_2) \theta_{XX} - 40q\newchi \theta_X
+ 220q \theta_X \theta_{XX}, \label{LW_eqn1} \\
\dot{\theta} & = & \frac{1}{5} \newchi + \theta_{XX}. \label{LW_eqn2}
\ea
Non-modulated $M_+$ states correspond to $\newchi=0$,
$\theta=\theta_0$ as there is a circle of equivalent $M_+$ states
related to each other by spatial translations. We consider the state
$\theta_0=0$ without loss of generality. We now use~(\ref{theta_eqn_lw})
and~(\ref{chi_eqn_lw}) to examine the two distinct linear instabilities
of the state $\newchi=\theta=0$ that are possible. Substituting
$\newchi=\hat{\newchi}(T)\e^{\i\ell X}$ and
$\theta=\hat{\theta}(T)\e^{\i\ell X}$ into~(\ref{theta_eqn_lw})
and~(\ref{chi_eqn_lw}) and linearising we obtain the Jacobian matrix
\ba
J_{\newchi,\theta} & = & \left( \begin{array}{cc} 
\xi_0 - \ell^2(1 + \xi_4) & -\ell^2 \xi_1 \\
1/a_2 & -\ell^2 \\
\end{array} \right), \nn
\ea
which has trace and determinant
\ba
\mathrm{tr_J}(\ell^2) & = & \xi_0-\ell^2(2+\xi_4), \nn \\
\mathrm{det_J}(\ell^2) & = & \ell^4(1+\xi_4) + \ell^2(\xi_1/a_2-\xi_0). \nn
\ea
Hence $\mathrm{det_J}'(0)=\xi_1/a_2-\xi_0$. The bifurcation to TW
occurs when $\xi_0=0$; it is the initial instability of $M_+$ when
$\xi_1/a_2>0$, i.e. (for the illustrative coefficient set) when
$20\hat{\mu}_1 - 3\hat{\mu}_2 = 0$ and
$6\hat{\mu}_1 - \hat{\mu}_2 > 0$. Similarly, the
LW instability occurs when $\xi_1/a_2=\xi_0$; it occurs before the TW
instability if $\xi_0<0$. These
conditions become $(20+1500q^2)\hat{\mu}_1=(3+250q^2)\hat{\mu}_2$ and
$20\hat{\mu}_1<3\hat{\mu}_2$ for our coefficient set. These lines are
sketched in figure~\ref{point_X_fig} and clearly correspond to the
behaviour of the TW and LW curves near $\X$ in
figure~\ref{mu1mu2_fig}(a).

The form of the equations~(\ref{LW_eqn1}) - (\ref{LW_eqn2}) stems
directly from the requirement of invariance under
$x$-reflection~(\ref{reflection_sym}),
which sends $(\newchi,\theta,\p_X) \ra (-\newchi,-\theta,-\p_X)$. We note
that the TW
bifurcation is one of the secondary instabilities of spatially
periodic patterns classified on symmetry grounds by Coullet \& Iooss
\cite{CI90}; their equations~(8a,b) closely resemble~(\ref{LW_eqn1}) -
(\ref{LW_eqn2}), although theirs contain only one bifurcation
parameter since they are concerned with classifying codimension-one
instabilities. By including a second bifurcation parameter,
$\kappa_2$, we are able to capture the codimension-two transition
between the TW and LW instabilities.

\subsubsection{Eckhaus instability dynamics near the LW bifurcation}

At the codimension-one LW
bifurcation, instability to modes of arbitrarily long wavelength
occurs; near this bifurcation we may adiabatically eliminate
$\newchi$ to derive a single real equation describing the dynamics. The
relevant scalings are
\ba
\theta = \epsilon \tilde{\theta}, \ \ \ \tilde{X}=\epsilon X, \nn \\
\newchi = \epsilon^3 \tilde{\newchi},   \ \ \ \tilde{T}=\epsilon^4 T, \nn
\ea
we introduce a bifurcation parameter $\nu$ by writing $\epsilon^2 \nu
= 1-\xi_1/(a_2 \xi_0)$. On substituting these scalings
into~(\ref{chi_eqn_lw}) we obtain
\ba
\xi_0 \newchi & = & -\xi_1 \theta_{XX} + \epsilon^2 \left[
\frac{(1+\xi_4)\xi_1}{\xi_0}\theta_{XXXX} +\left( \frac{\xi_1
\xi_2}{\xi_0}-\xi_3\right) \theta_X \theta_{XX} \right] +
O(\epsilon^4) \nn
\ea
and we substitute this expression for $\newchi$ into the $\dot{\theta}$
equation~(\ref{theta_eqn_lw}) to obtain
\ba
\dot{\theta} & = & \nu \theta_{XX} + \frac{(1+\xi_4)\xi_1}{a_2
\xi_0^2}\theta_{XXXX} + \frac{\xi_1 \xi_2-\xi_0 \xi_3}{a_2 \xi_0^2} \theta_X
\theta_{XX}, \label{LW_eqn3}
\ea
to leading order. Equation~(\ref{LW_eqn3}) is identical in form to
that which describes the weakly nonlinear behaviour of the Eckhaus
instability. Like the Eckhaus instability, the LW bifurcation is
therefore always subcritical.

\subsubsection{Instability of travelling waves near the TW bifurcation}

In a similar way, the dynamics of travelling waves, near the TW
bifurcation, near $\X$, can be investigated.
At the TW bifurcation, $\theta$
can be eliminated (again, adiabatically) from~(\ref{LW_eqn1}) -
(\ref{LW_eqn2}) and the resulting single real Ginzburg--Landau
equation for $\newchi$ describes the bifurcation leading to TW
solutions. The scalings leading to~(\ref{theta_eqn_lw})
and~(\ref{chi_eqn_lw}) cannot be chosen to include a term $\newchi^3$
in~(\ref{chi_eqn_lw}) which would be required to capture stable
finite-amplitude TW states near the bifurcation point. In the absence
of modulations a different set of scalings can be chosen which are
able to include this term. In this way the sub- or supercriticality of
the TW bifurcation can be easily computed. 

However, it transpires that spatially-periodic TW are unstable to
modulational disturbances and this instability is indeed captured by the
reduced equations near $\X$. To illustrate this, we compute the stability of
the $X$-independent state $\newchi=\newchi_0$ constant, $\theta=\newchi_0
T/a_2$ for~(\ref{theta_eqn_lw}) and~(\ref{chi_eqn_lw}).

Let
\ba
\newchi = \newchi_0(1+\alpha \e^{\i\ell X} + c.c.), & \ \ \
\theta = \frac{\newchi_0}{a_2}(T+\beta \e^{\i\ell X} + c.c.), \nn
\ea
then, on substituting into~(\ref{theta_eqn_lw}) and~(\ref{chi_eqn_lw})
and linearising we obtain
\ba
\dot{\alpha} & = & \xi_0 + \alpha [\xi_0-\ell^2(1+\xi_4)] +
\frac{\beta}{a_2}(\i\ell \xi_2 \newchi_0 - \ell^2 \xi_1), \nn \\
\dot{\beta} & = & \alpha - \ell^2 \beta. \nn
\ea
Eliminating $\alpha$ yields a single linear, constant coefficient ODE
for $\beta$:
\ba
\ddot{\beta} + \dot{\beta}(\ell^2(1+\xi_4)-\xi_0) + \beta (\ell^2(1 -
\xi_0 + \xi_1/a_2 + \ell^2(1+\xi_4)) - \i\ell \xi_2
\newchi_0/a_2) & = & \xi_0. \nn
\ea
Solving the homogeneous equation for the complementary functions
$\beta=\e^{\lambda_\pm T}$ we find
\ba
\lambda_\pm & = & \frac{\xi_0 -\ell^2(1+\xi_4) \pm
\sqrt{(\xi_0 - \ell^2(1+\xi_4))^2 - 4(\ell^2(1 -
\xi_0 + \xi_1/a_2 + \ell^2(1+\xi_4)) - \i\ell \xi_2
\newchi_0/a_2))}}{2}. \nn
\ea
Expanding this expression for $\lambda_+$ up to $O(\ell^2)$ we obtain
\ba
\mathit{Re}(\lambda_+) & = & \xi_0 +\ell^2\left( \frac{\xi_2^2
\newchi_0^2}{16 a_2^2 \xi_0^3} -
\frac{1}{\xi_0} - \frac{\xi_1}{a_2 \xi_0} - \xi_4 \right) + O(\ell^4). \nn
\ea
For sufficiently small $\xi_0$, the growth rate $\lambda_+$ is more
positive than the growth rate $\xi_0$ of the state
$\newchi=\newchi_0$, $\theta=\newchi_0 T/a_2$ towards a
fully-nonlinear TW equilibrium, hence stable TW are not anticipated to
appear close to the codimension-two point $\X$.

\begin{figure}
\begin{center}
\includegraphics[width=12.0cm]{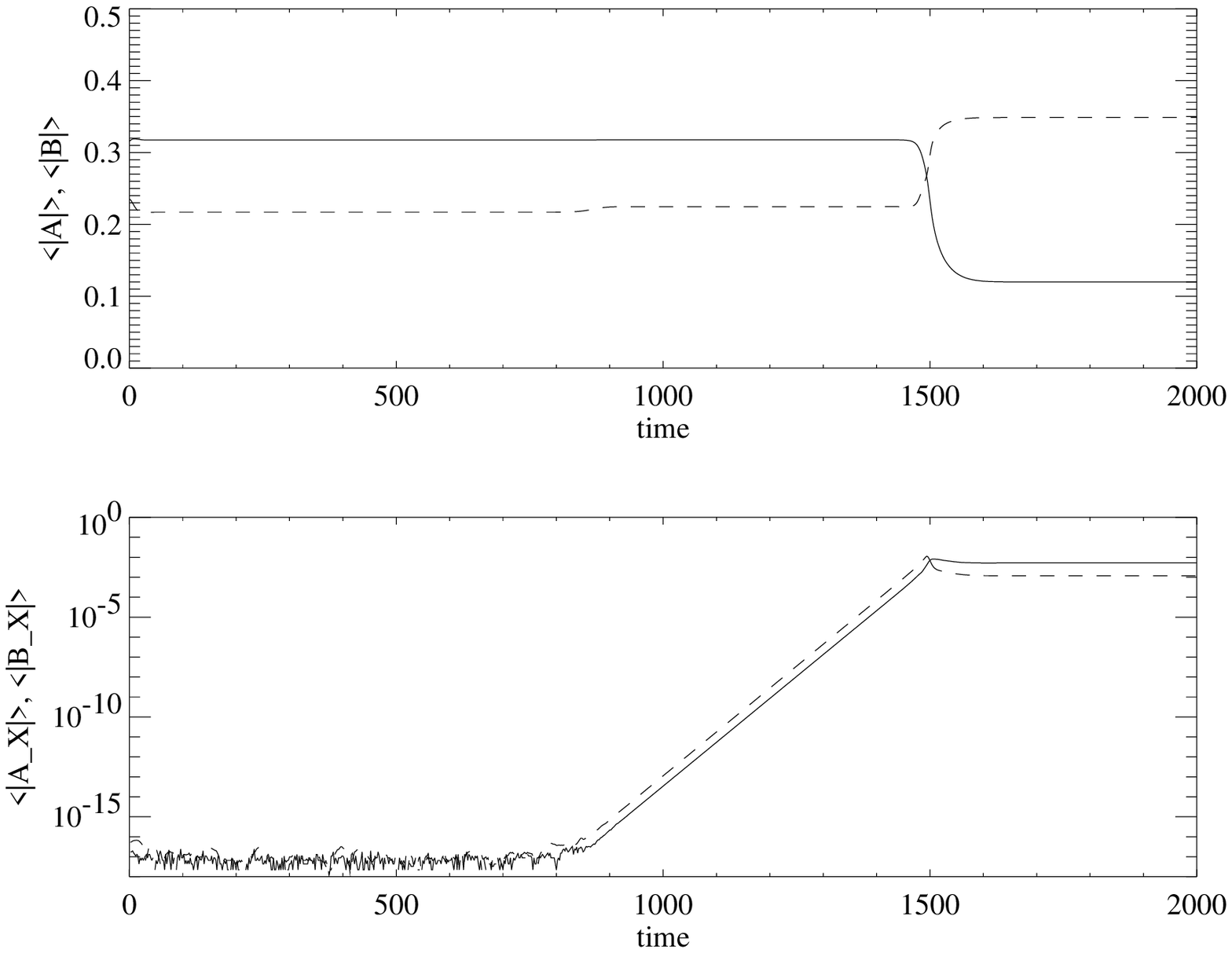}
\includegraphics[width=12.0cm]{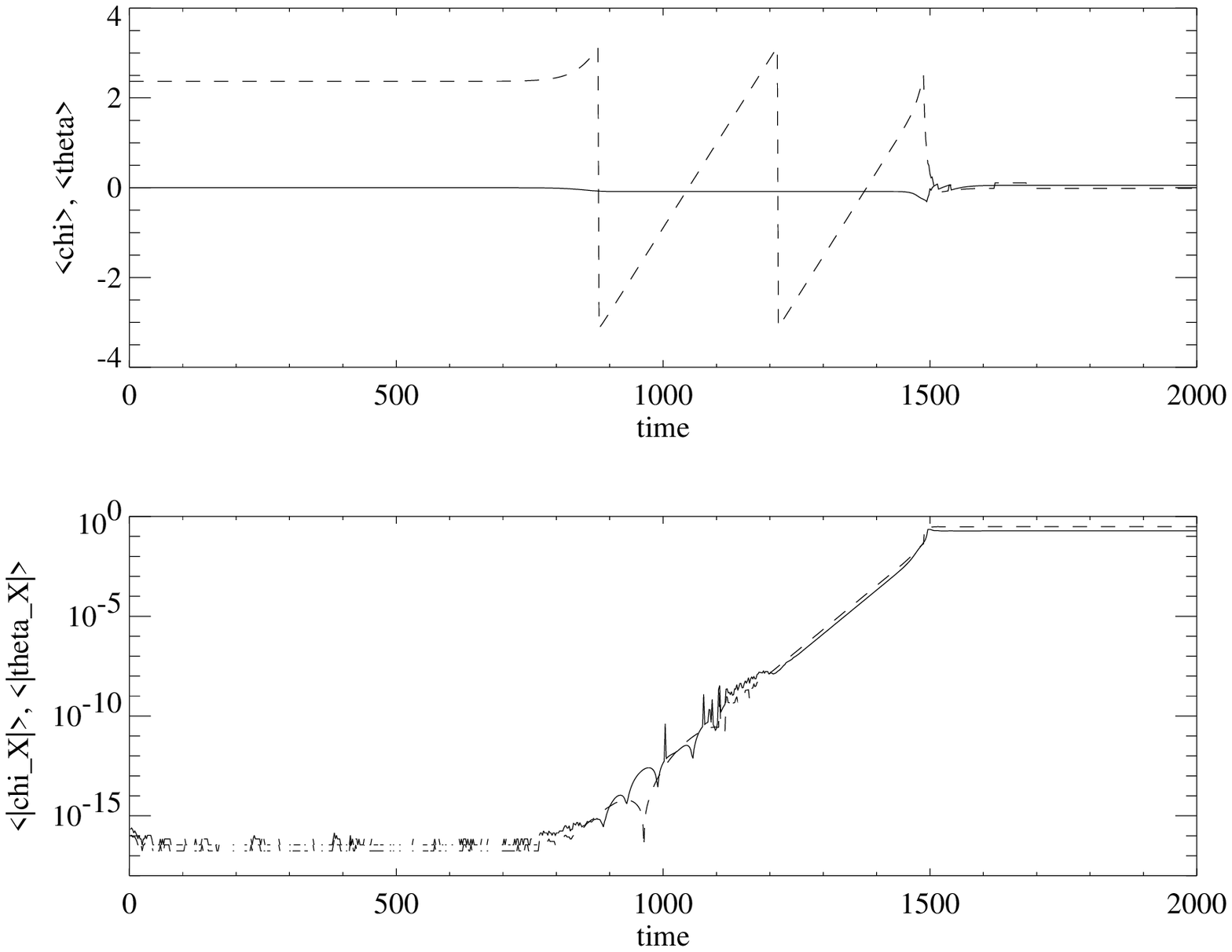}
\caption{Time evolution of perturbation from the $M_+$
equilibrium near $\X$. $\mu_1=0.018$, $\mu_2=0.7$; all other coefficients
are as in figure~\ref{mu1mu2_fig}.The four panels show spatial
averages of: (a) $|A|$ (solid
line) and $|B|$ (dashed line), (b) $|A_X|$ (solid) $|B_X|$ (dashed),
(c) $\newchi$ (solid), $\theta$ (dashed), (d) $|\newchi_X|$ (solid),
$|\theta_X|$ (dashed). An initial transient growth
of a TW mode (illustrated by the constant rate of evolution of
$\theta$ for $900<t<1500$) gives way to a spatially-modulated steady
state at $t=1600$.}
\label{m_plus_instability_fig}
\end{center}
\end{figure}

\begin{figure}
\begin{center}
\includegraphics[width=12.0cm]{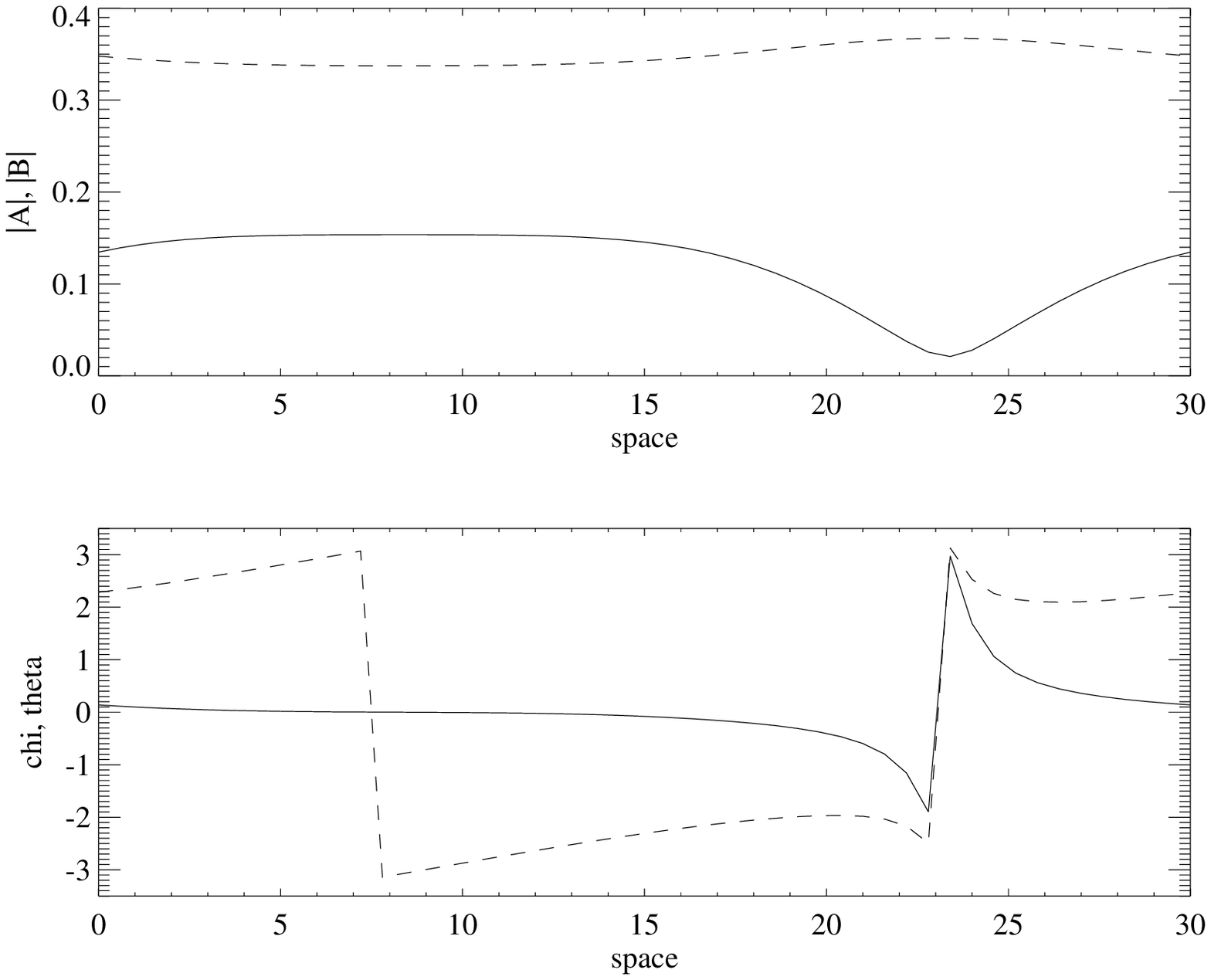}
\caption{Steady spatially-modulated state reached at the end of the
numerical integration illustrated in
figure~\ref{m_plus_instability_fig}. $\mu_1=0.018$, $\mu_2=0.7$; all
other coefficients are as in figure~\ref{mu1mu2_fig}. (a) $|A|$
(solid), $|B|$ (dashed). (b) $\newchi$ (solid), $\theta$ (dashed).}
\label{m_plus_instability002_fig}
\end{center}
\end{figure}

Near $\X$, numerical results (see figures~\ref{m_plus_instability_fig}
and~\ref{m_plus_instability002_fig})
show $M_+$ solutions losing stability first to
a TW perturbation which generates a TW-like `transient', and then the
occurrence of an instability to spatial modulations.
The final state is, however, steady, and consists of an extremely
long-wavelength, but $O(1)$ in amplitude, modulation - see
figure~\ref{m_plus_instability002_fig}. Throughout most
of the domain $\theta$ and $\phi$ increase linearly with $X$, while
$\newchi$ remains close to zero.

From the form of the equations~(\ref{eqn_R}) - (\ref{eqn_phi}) it is
clear that families of solutions with $\theta_X$ and $\phi_X$
constant and non-zero are possible, corresponding to exactly
spatially-periodic states with wavenumber close to unity. Some
analytic investigation of these solutions should be possible.
It may well be possible also to look analytically at other classes of
solution, for example homoclinic orbits corresponding to spatially
localised structures in infinite domains, or
solutions where $R$ and $S$ are spatially periodic, but the
possibilities are too numerous to discuss further here.

\subsection{Dynamics near the codimension-two point $\Y$}

Below the point $\A$ on figure~\ref{mu1mu2_fig} the LW and SW curves
intersect at yet another
codimension-two point, labelled $\Y$ on figure~\ref{mu1mu2_fig}(a). A
similar analysis to that near $\X$ could be carried out here, leading to
a pair of equations
for the SW instability (which leads to time-periodic variations in
solution amplitude), and the LW (phase) instability. Although we have
not computed the reduced `normal form' in this case, its structure is
simple to derive and we include it for completeness. The relevant
reduced equations describing the dynamics of
this bifurcation are coupled ODEs for two variables $(z,\theta)$ where
$z \in
\mathbb{C}$ gives the perturbation in the direction of the SW
instability and $\theta \in \mathbb{R}$ describes the LW instability.
On symmetry grounds, the equations will be invariant under the
transformation $(z,\theta,\p_X) \ra (z,-\theta,-\p_X)$. We also
make use of the normal form
symmetry $z \ra \e^{\i\psi} z$ which appears naturally in Hopf
bifurcation problems, and we use the fact that only spatial
derivatives of $\theta$ will appear because the value of $\theta$
itself is dynamically unimportant. Under these constraints the reduced
equations (including terms up to cubic order in $(z,\theta,\p_X)$)
take the form
\ba
\dot{z} & = & (\lambda_1 + \i\omega)z - \xi_0 z|z|^2 + \xi_1 z\theta_X
+ \xi_2 z_{XX}, 
\label{y_eqn1} \\
\dot{\theta} & = & - \lambda_2\theta_{XX} - \xi_3 \theta_{XXXX}
+\i\xi_4 \left( z_X\bar{z} - \bar{z}_X z \right) +\xi_5 |z|^2_X, \label{y_eqn2}
\ea
where $\xi_0,\xi_1,\xi_2 \in \mathbb{C}$ and $\xi_3,\xi_4,\xi_5 \in
\mathbb{R}$ are undetermined coefficients, $\omega>0$
is the frequency of oscillation at the Hopf bifurcation, and
$\lambda_1$, $\lambda_2$ are real bifurcation parameters. 
Due to the large number of
undetermined coefficients, space does not permit a detailed
investigation of this bifurcation here. However, since the symmetry
does not permit a linear term in $z$ in the $\dot{\theta}$ equation it
is clear that the dynamics of~(\ref{y_eqn1}) - (\ref{y_eqn2}) are
not related to those of~(\ref{theta_eqn_lw})
and~(\ref{chi_eqn_lw}). The codimension-one bifurcation that occurs for
$\lambda_1=0$ and $\lambda_2<0$ was one of the `normal
forms' identified by Coullet \& Iooss \cite{CI90} and has been
explored numerically by Lega \cite{L91} and Daviaud et
al. \cite{DLBCD92}. In particular, these authors identify two distinct
regimes of spatio-temporal chaos depending on the choices of the
coupling coefficients in~(\ref{y_eqn1}) - (\ref{y_eqn2}).

\subsection{Instability of $M_-$ to long-wavelength perturbations}

Finally in this section, we briefly discuss the dynamics in the
quadrant $\mu_1>0$, $\mu_2<0$. The spatially-periodic equilibrium
state $M_{-}$ exists in the whole of this region and is stable when
$\mu_2$ is sufficiently negative, for a fixed $\mu_1>0$. The
amplitudes $R=R_0$ and $S=S_0$ satisfy
\ba
0 & = & \mu_1 -q^2 - S_0 - a_1 R_0^2 - b_1 S_0^2, \label{R0_eqn002} \\
0 & = & \mu_2 S_0 + R_0^2 - a_2 S_0^3 - b_2 R_0^2 S_0, \label{S0_eqn002}
\ea
and $\newchi=\pi$. For $q=0$, $M_-$ states lose stability
to TW solutions as $\mu_2$ is increased at fixed positive
$\mu_1$. These TW states have $\newchi \approx \pi$ near the bifurcation
since $\newchi=\pi$ for $M_-$. As for $M_+$ this phase instability
also generates a distinct long-wavelength instability when $q \neq 0$.

\begin{figure}
\begin{center}
\includegraphics[width=12.0cm]{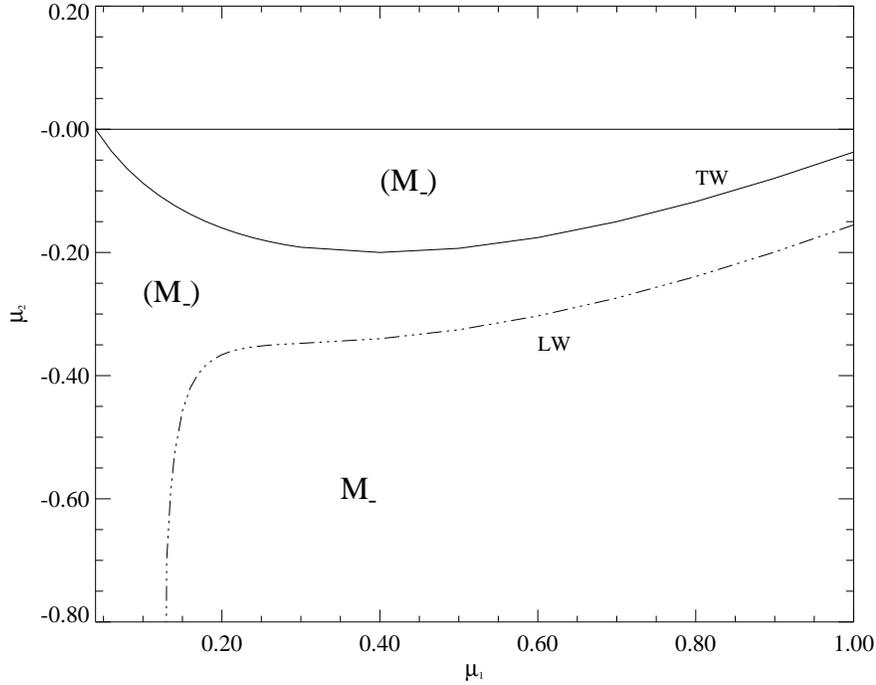}
\caption{Region of stability of the mixed mode $M_-$ for the
illustrative set of coefficients $a_1=1$, $a_2=5$, $b_1=2$, $b_2=0$,
$c=1$, $q=0.2$. The region of stable $M_-$ is bounded by TW (solid
line) and LW (dash-dotted line) instabilities. Note the asymptotic
behaviours of the LW curve: at large negative $\mu_2$ we find $\mu_1
\sim 3q^2 =0.12$ for this set of coefficients; at large $\mu_1$ it
closely follows the TW curve.}
\label{m_minus_fig}
\end{center}
\end{figure}

Following the usual linearisation of~(\ref{eqn_R}) - (\ref{eqn_phi})
about $M_-$ we find that a long-wavelength instability occurs when
\ba
(2c S_0^2 - R_0^2)(2S_0^3(b_1b_2-a_1a_2)+(b_2-2b_1)S_0^2-S_0-a_1R_0^2)
& = & 2q^2(R_0^2+2a_2S_0^3). \nn \\ & & \label{m_minus_lw}
\ea
Since the bifurcation to TW occurs when $R_0^2=2S_0^2$ as before, it
is clear that in the case $c=1$ these bifurcation curves do not
intersect and no further codimension-two bifurcation points appear. 
In the limit where we keep $\mu_1 \sim O(1)$ fixed and let
$\mu_2 \ra -\infty$ we see from~(\ref{R0_eqn002}) - 
(\ref{S0_eqn002}) that $S_0 \ra 0$ and $R_0 \sim
O(1)$. Then~(\ref{m_minus_lw}) is satisfied when $a_1 R_0^2 \approx
2q^2$ to leading order. From~(\ref{R0_eqn002}) this gives the asymptotic
behaviour of the curve of LW instability: $\mu_1 \sim 3q^2$, 
illustrated in figure~\ref{m_minus_fig}.

In summary, in this section we have identified various codimension-one
and two bifurcations involving modulational instabilities. In
particular, two codimension-2
points, $\X$ and $\Y$, organise the interaction of the new LW
instability with the previously-studied TW and SW
bifurcations. Reductions of the original PDEs~(\ref{eqn1}) -
(\ref{eqn2}) to `normal form' equations describing the dynamics near
these bifurcations enables us to gain insight into the nature of these
bifurcations, despite the fact that solutions, even of the reduced
equations, may have extremely complicated spatial structure.


\section{The heteroclinic cycle}

\label{sec:cycle}

One of the most interesting features of the ODE problem~(\ref{eqn03})
- (\ref{eqn05}) analysed in
Part 1 is the existence of a robust heteroclinic cycle between pure
mode solutions related by a half-wavelength spatial translation;
for example $P_+$ and $P_-$. For the coefficient values selected in this
paper this heteroclinic cycle exists for an open
region of the $(\mu_1,\mu_2)$ plane, abutting the origin. Its formation
relies on the existence of pairs of two-dimensional invariant subspaces for the
dynamics: for example within the two-dimensional subspace
$\mathrm{Fix}(m_x)=\{Im(A)=Im(B)=0\}$ the point $P_+$ is a saddle and
$P_-$ is a sink
and, if various other conditions are met, a connecting trajectory
between them exists. The second connection is then forced to exist
by symmetry, and is contained within the subspace
$\mathrm{Fix}(m_x \circ \tau_\pi)=\{Re(A)=Im(B)=0\}$. 

For small positive $\mu_2$ and increasing $\mu_1$ the cycle is formed
after a global bifurcation that involves 
the intersection of the unstable manifold of $P_+$ and the stable
manifold of the origin; this bifurcation also creates or destroys the
SW periodic orbit. At larger values of $\mu_2$ the connecting
trajectory appears after a saddle-node bifurcation marking the
boundary of existence of a further two $M_+$ equilibria. Hence this
curve of saddle-node bifurcations also bounds the region of
existence of the cycle. At larger $\mu_1$ the cycle ceases to exist 
where the pure mode equilibria $P_\pm$ gain stability in a
pitchfork bifurcation with $M_-$ equilibria. These bifurcations
proscribe the region of existence of the cycle. 
Within this region of existence, a further curve separates
regions where it is stable or unstable. At this stability boundary a
branch of modulated waves (MW) bifurcate from the cycle in a global
(`resonant') bifurcation. Theoretical work by Armbruster et
al. \cite{AGH87} and by Proctor \& Jones \cite{PJ88} was confirmed by
the general stability results of Krupa \& Melbourne \cite[section
6.1]{KM95}. It turns out that the natural condition (that the ratio
of eigenvalues in the `contracting' and `expanding' directions should
be greater than one for stability) is necessary and sufficient. This yields
the condition $\mu_1<b_1\mu_2/a_2$ for the stability boundary of the cycle.

When modulational terms are included, the instability of $P_+$
to $P_-$ that is necessary for existence of the cycle implies that
$P_+$ will also be unstable to sufficiently
long-wavelength perturbations, so it might be expected that the cycle
also could not be stable to long-wavelength perturbations. Moreover,
the subspaces $\{Im(A)=Im(B)=0\}$ and
$\{Re(A)=Im(B)=0\}$ are no longer invariant for
the dynamics; spatial variations of the amplitude $A$ drive the
evolution of the phase $\theta$ when $q \neq 0$, see
equation~(\ref{eqn_theta}). Of course, the cycle still exists, within
the subspace of solutions with no spatial modulation at all. But, in
large domains it cannot be asymptotically stable (that is, points in a
full neighbourhood  - in some appropriate sense - of the cycle
converge to it) as the equilibria $P_\pm$ will be unstable to
perturbations of a sufficiently long wavelength, as well as to each
other within the subspaces containing the heteroclinic connecting
orbits.

Numerical simulations show that spatially-modulated perturbations
eventually grow, when $q \neq 0$, and the simulation converges to a
periodic orbit rather than showing the characteristic increases of
time spent near each equilibrium that would indicate convergence to the
robust heteroclinic cycle.

In this section we analyse the behaviour of trajectories close to the
cycle using
a linearised stability analysis near the equilibria $P_\pm$ and under
the commonly-used assumption that trajectories near the cycle spend
very little time passing between neighbourhoods of the equilibria.
Our aim is to determine conditions for existence of periodic orbits
lying close to the cycle. We restrict our attention to periodic orbits
that spend equal amounts of time near each equilibrium since this
feature is observed in numerical work.


Recall from section~\ref{sec:pure_mode} that the linear stability of
$P_+$ to perturbations $\sim \alpha_1\e^{\i\ell X} +
\bar{\alpha}_2\e^{-\i\ell X} $ in $A(X,T)$ with wavenumber $\ell$ is given by
the matrix~(\ref{pplus_linear}) acting on the amplitudes
$(\alpha_1,\alpha_2)$. For simplicity we write this matrix as
\ba
M_1 & = & \left( \begin{array}{cc} 
a_0 - \hat{q} 	& b_0 \\
b_0 		& a_0 + \hat{q} \\
\end{array} \right), \nn
\ea
where
\ba
a_0 & = & \mu_1-q^2-b_1 B_0^2-\ell^2, \label{defn_a} \\
b_0 & = & B_0, \\
\hat{q} & = & 2q\ell.
\ea
Note that a necessary condition for the existence of the cycle is that
the eigenvalues $\lambda_\pm$ of $M_1$, when evaluated when $\ell=0$,
satisfy $\lambda_- < 0 < \lambda_+$, so that $P_+$ is a saddle
point. The form of $M_1$
implies that this is equivalent to requiring $a_0<0$ when $\ell=0$, and
hence in the rest of this section we assume $a_0<0$. The
corresponding linearisation around $P_-$ is denoted $M_2$:
\ba
M_2 & = & \left( \begin{array}{cc} 
a_0 - \hat{q} 	& -b_0 \\
-b_0 		& a_0 + \hat{q} \\
\end{array} \right). \nn
\ea
The form of $M_2$ is determined entirely from $M_1$ and the
equivariance of the ODEs~(\ref{eqn1}) - (\ref{eqn2}).

We assume that trajectories close to the heteroclinic cycle spend
equal amounts of time $T$ near each equilibrium, and that we may
ignore the time spent travelling between neighbourhoods of the
equilibria. Hence a perturbation $v_0=(\alpha_1,\alpha_2)^T$ near
$P_+$ evolves to the point
\ba
v_1 & = & \calS v_0 = \exp(M_2 T) \exp(M_1 T) v_0, \nn
\ea
where the matrix $\calS$ is given by
\ba
\calS & = & 
\frac{1}{2c_0^2}
\left(
\begin{array}{cc}
2b_0^2 e_+ e_- + \hat{q}e_+^2(q-c_0) + q e_-^2(q+c_0) & -b_0q(e_+ - e_-)^2 \\
b_0q(e_+-e_-)^2					& 2b_0^2 e_+ e_- + q
e_+^2(q+c_0) + q e_-^2 (q-c_0)
\end{array}
\right), \nn
\ea
and $c_0=\sqrt{\hat{q}^2 + b_0^2}$ and $e_\pm=\exp[(a_0 \pm c_0)T]$. The
eigenvalues of $\calS$ then determine the stability of the
cycle. We are particularly interested in the dependence of these
eigenvalues on the travel time $T$, the selected wavelength $\ell$ and the
wavenumber mismatch parameter $q$.

When $q=0$, $\calS$ simplifies enormously: it is diagonal with
eigenvalues $e_+ e_-=\exp(2a_0T)$ where $a_0$ is defined
in~(\ref{defn_a}). Since $a_0<0$ we have contraction of the perturbation
vector $v_0$ under successive iterates of $\calS$. This corresponds to
trajectories passing repeatedly through neighbourhoods of $P_\pm$, and
converging to the cycle. This result is independent of the time $T$ spent
near each equilibrium.

For general $q \neq 0$ the eigenvalues of $\calS$ will, however, depend
on $T$. Moreover, for any $q \neq 0$, $\calS$ will have eigenvalues of
modulus greater than unity when $T$ is taken to be sufficiently large
(so that $e_+$ is sufficiently large). So, in addition to the
discussion above, we can conclude directly from the form of the map
that for any non-zero $q$ the cycle is unstable.

\begin{figure}[!h]
\begin{center}
\includegraphics[width=14.0cm]{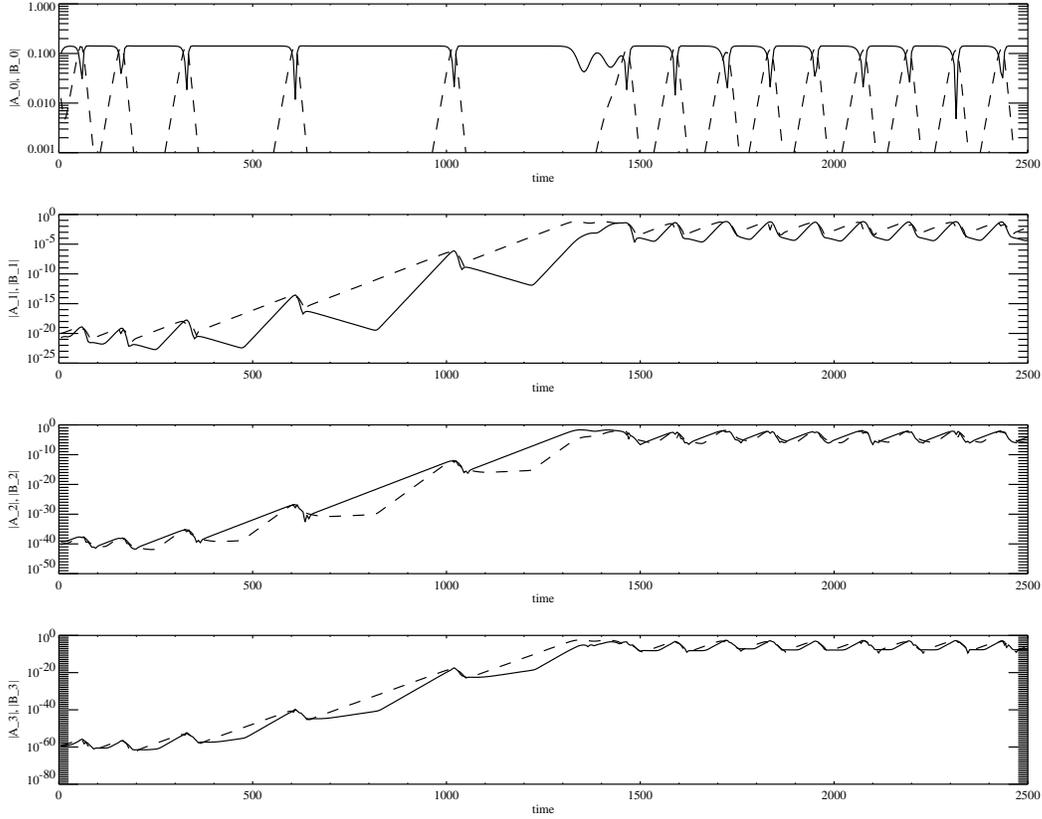}
\caption{Temporal evolution of the first few Fourier modes (writing
$A(X,T)=\sum_{j=0}^N \hat{A}_j \e^{2\pi \i j/L}$ and $B(X,T)$
similarly) in a solution starting near the subspace of non-modulated
solutions and converging to the periodic orbit. The top graph shows
the time evolution of $|\hat{A}_0|$ (dashed) and $|\hat{B}_0|$
(solid). The other three graphs show $|\hat{A}_j|$ (dashed) and
$|\hat{B}_j|$ (solid) for $j=1,2,3$ respectively.Note the initial
decay when the travel time $T$ is short, followed by growth of the
perturbation as $T$ increases.Parameter values are $L=30$,
$\mu_1=10^{-6}$, $\mu_2=0.1$, $q=10^{-3}$, and the coefficients are as
in figure~\ref{mu1mu2_fig}.}
\label{close_to_cycle}
\end{center}
\end{figure}

Numerical simulations indicate that trajectories remain close to the
cycle, though, so it is natural to ask whether stable dynamics close
to the cycle is possible. One possibility is the existence of a
long-period periodic orbit. That such an orbit might exist is
motivated by the observation that $\calS$ has an eigenvalue greater
than unity for $T \gg 1$, i.e. trajectories very close to the cycle
are pushed further away from it, but that for $T \sim 1$ the
eigenvalues of $\calS$ may lie within the unit circle, indicating that
trajectories that start further away from the cycle move closer on
successive passes near the cycle. This behaviour is confirmed by
numerical simulations (see figure~\ref{close_to_cycle}), where we
choose an initial condition lying extremely close to the subspace of
non-modulated solutions, but not close to the heteroclinic cycle.
Thus
the first few travel times $T$ between neighbourhoods of $P_\pm$ occur
with $T$ small, and the perturbation decays. Then, as the trajectory
becomes closer to the cycle $T$ increases and the perturbation grows
(exponentially on average) until the trajectory converges to the
periodic orbit.

The above argument is unusual in that it suggests that we can extract
nonlinear information (about the period of the orbit) from a purely
linear calculation (that which leads to the form of $\calS$). This is
because we are essentially using $T$ as a proxy for the closest
distance between the periodic orbit and the heteroclinic cycle. Hence
solving for the position of the orbit and solving for $T$ are really
the same thing. By setting up the usual `small box' approach we could
construct a return map, fixed points of which would correspond to periodic
orbits. By using $T$ instead, we are able to circumvent the need to
compute this return map in full in order to extract an estimate for
the period. Essentially we construct the component of the return map
in a pair of directions orthogonal to the invariant plane corresponding to
unmodulated solutions. Then the condition for locating a fixed point
of this part of the map is identical to the condition that $\calS$ has
an eigenvalue of $+1$.

At least for small $q$ and large times $T$ we expect that the
behaviour outlined above could be captured by the map $\calS$ based on
linearisation near $P_\pm$. We compute the approximate period $P(q)
\approx 2T(q)$ of a stable periodic orbit by
imposing the condition that the larger eigenvalue $m_+$ of $\calS$ is $+1$
(the other ($m_-$) lies inside the unit circle). Finally we consider
only the minimum wavenumber $\ell_{min}=2\pi/L$ since this is the mode
with the highest growth rate for the points in the $(\mu_1,\mu_2)$
plane that we consider. Hence
the eigenvalue condition yields a relationship (implicitly) between $T$ and $q$
which can be compared with numerical simulations. This relationship
simplifies in the asymptotic regime where $q$ is small and $T$ is large.

\begin{figure}
\begin{center}
\includegraphics[width=13.0cm]{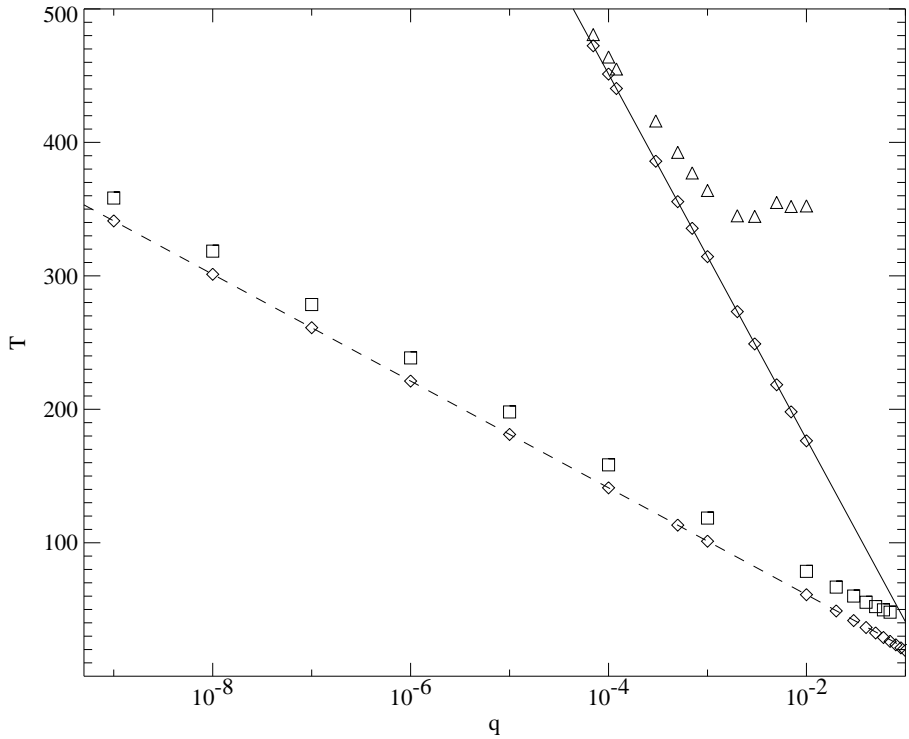}
\caption{The dependence on $q$ of the approximate half-period $T(q)$
of the periodic orbit near the heteroclinic cycle, for two
combinations of $L$ and $\mu_2$. In both cases, the illustrative
coefficient values were used, as for
figure~\ref{mu1mu2_fig}, and $\mu_1=q^2$. Squares
($\Box$) are the results
of numerical simulations for $L=30$ and $\mu_2=0.1$, and triangles
($\triangle$) are numerical results for $L=25$, $\mu_2=0.05$. 
Diamonds $(\Diamond$) give the exact $T$--$q$
relationship implied by the linearised analysis near the cycle using
$\ell=\ell_{min}=2\pi/L$, in both cases, and the asymptotic result of
the linearised analysis is shown by the dashed and solid lines.}
\label{tq_fig}
\end{center}
\end{figure}

Substituting for $e_\pm=\exp[(a_0 \pm c_0)T]$ in $\calS$ and computing its
eigenvalues we find
\ba
m_\pm & = & \frac{\left[ 2b_0^2 \e^{2c_0T} + \hat{q}^2\e^{4c_0T} +
\hat{q}^2 \pm Y \right] \e^{2a_0T}}{2(b_0^2+\hat{q}^2)\e^{2c_0T}}, \label{exact_tq}
\ea
where
\ba
Y^2 & = & 4\hat{q}^2 b_0^2 \e^{6c_0T} + 4\hat{q}^2 b_0^2 \e^{2c_0T} +
\hat{q}^4 \e^{8c_0T} - 2\hat{q}^4 \e^{4c_0T} + \hat{q}^4 - 8\hat{q}^2 b_0^2
\e^{4c_0T}. \nn
\ea
It might appear that the dominant term for large $T$ is the first term
in the expression for $Y^2$, proportional to $\hat{q}^2 \e^{6c_0T}$,
but this is incorrect. In fact, the dominant terms in the limit of 
small $\hat{q}$ and large $T$ are those proportional to $\hat{q}^2
\e^{4c_0T}$. This leads to the asymptotic relationship
\ba
T & \sim & \frac{1}{a_0+b_0} \log \left( \frac{b_0}{\hat{q}}\right), \nn
\ea
as $q \ra 0$. For two combinations of $L$ and $\mu_2$,
figure~\ref{tq_fig} compares this asymptotic relationship with the exact
implicit $T$--$q$ relationship implied by~(\ref{exact_tq}) setting
$m_+=1$, and with the results of numerical integrations of the
PDEs~(\ref{eqn1}) - (\ref{eqn2}). In both cases there is clear
agreement as long as $q$ is sufficiently small. At small $q$ the
numerical simulations become more difficult, due to the intermittent
nature of the dynamics.  
Given the several approximations involved in the analytic estimate of
$T(q)$ the results of figure~\ref{tq_fig} are enouraging. The major
discrepancy in the analysis is the consideration of only one wavenumber
$\ell_{min}=2\pi/L$ in the estimate. Implicitly we assume that the `most
unstable' eigenfunction direction near $P_+$ is aligned exactly with
the single Fourier mode $\e^{\i\ell_{min} X}$.

\begin{figure}[!h]
\begin{center}
\includegraphics[width=12.0cm]{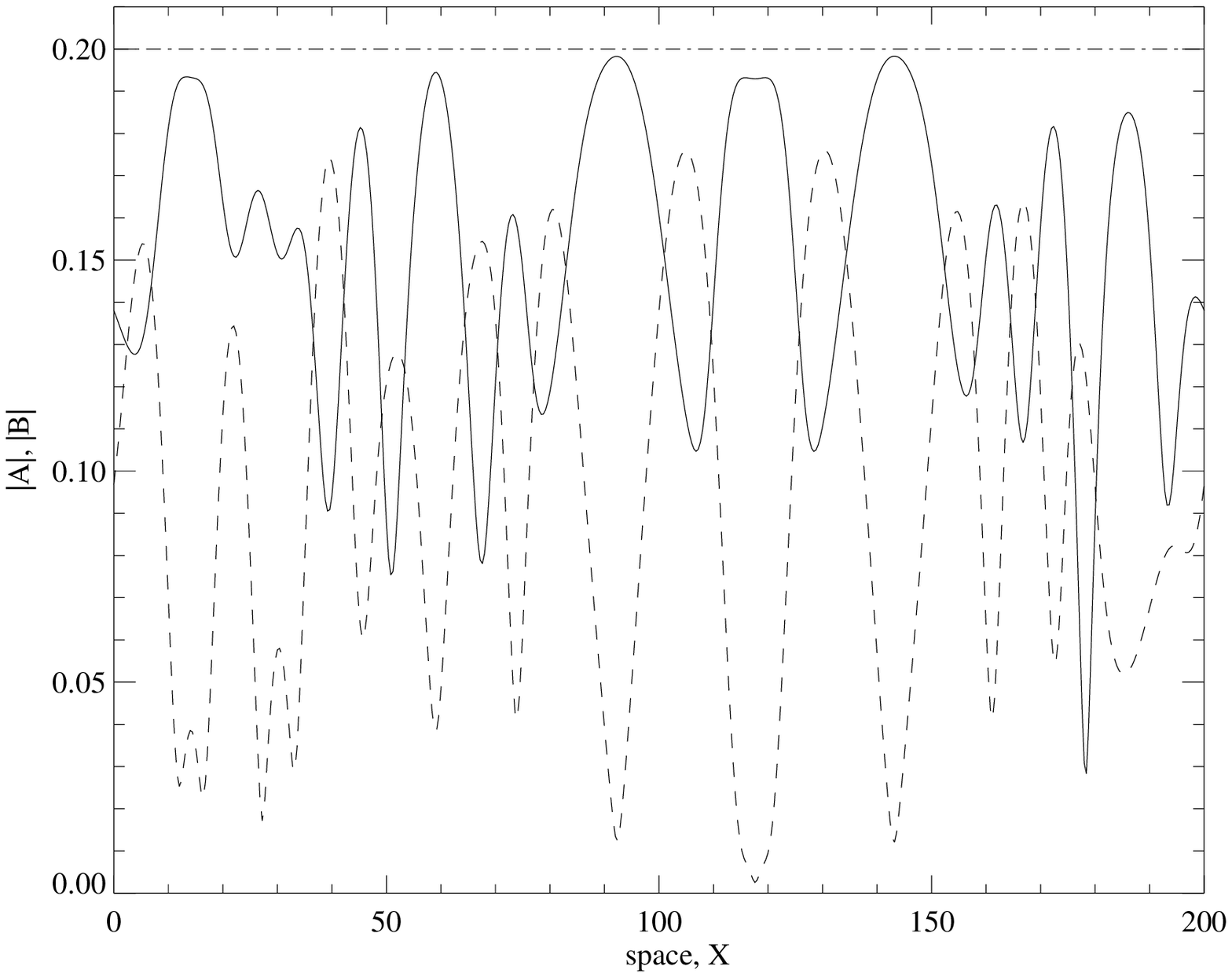}
\caption{The moduli $|A|$ (dashed line) and $|B|$ (solid line) for an 
instantaneous view of a typical solution of the PDEs~(\ref{eqn1}) -
(\ref{eqn2}). The horizontal dash-dotted line is at
$|B|=\sqrt(\mu_2/a_2)=0.2$ corresponding to the amplitude of the $P_+$ equilibrium. 
The parameter values are $q=0.2$, $\mu_1=0.04$
and $\mu_2=0.2$; the illustrative coefficient values are used and the
domain is of length $L=200$ with periodic boundary conditions.}
\label{het_fig2}
\end{center}
\end{figure}

\begin{figure}[!h]
\begin{center}
\includegraphics[angle=90,width=13.4cm]{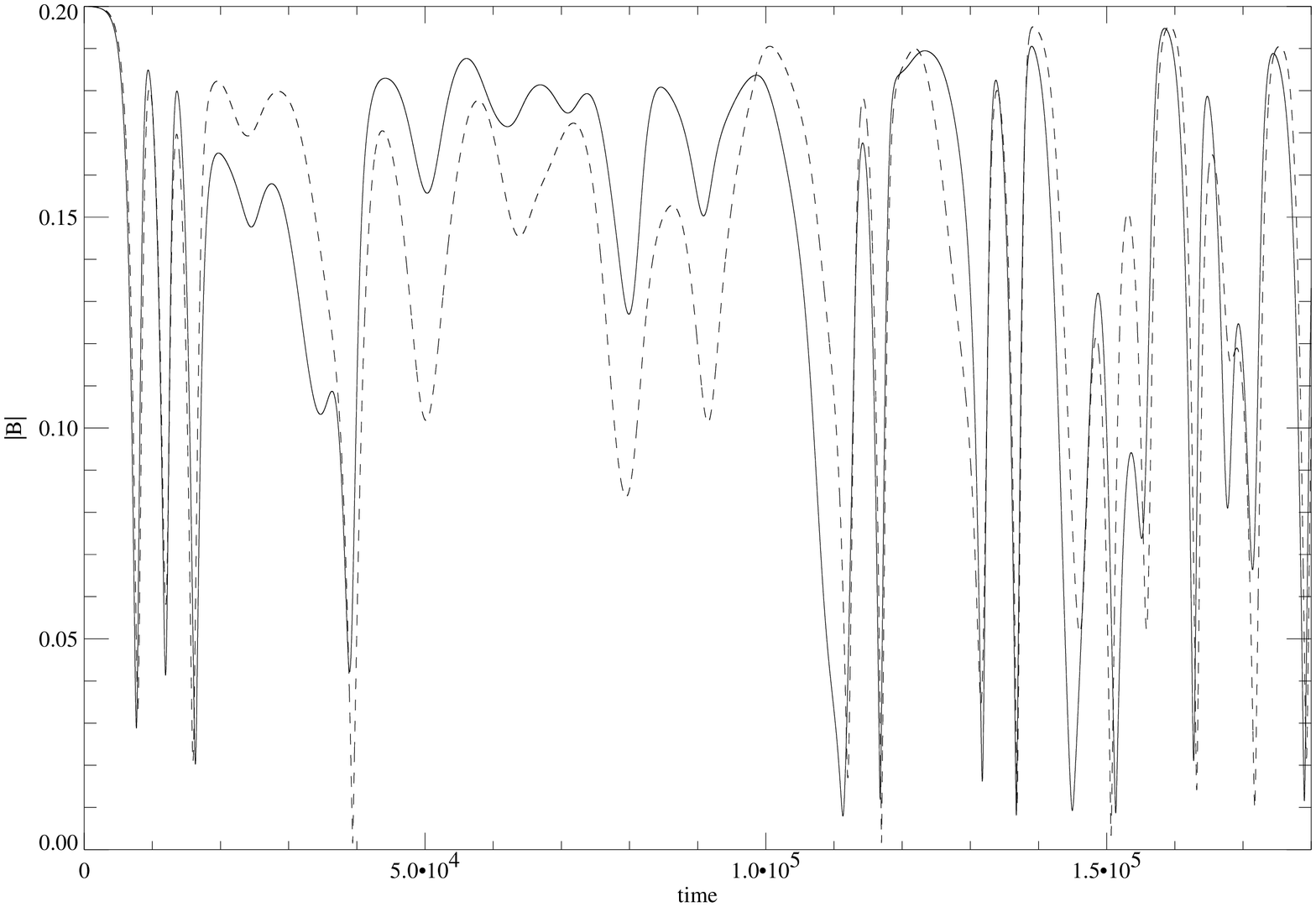}
\caption{Time evolution of $|B|$ at two spatial locations, $X=105$ (solid line) and $X=110$
(dashed line), for a typical solution of the PDEs~(\ref{eqn1}) -
(\ref{eqn2}). The parameter values, coefficients and domain size are as for
figure~\ref{het_fig2}.}
\label{het_fig}
\end{center}
\end{figure}

For larger $q$ numerical simulations did not converge to a stable
periodic orbit, but instead remained spatiotemporally disordered even
after large integration times. Although solutions still often spend
considerable amounts of time near the spatially-uniform $P_+$ state
these events occur only intermittently. At much larger $q$,
numerical integrations of the PDEs show that the cycle still plays a
r\^ole in \emph{local} organisation of the spatiotemporally complicated
dynamics. The spatial structure of
solutions often resembles a series of fronts between
intervals of points which remain near either $P_+$ or $P_-$ for a
time, and then switch rapidly to a neighbourhood of the other
equilibrium. This is illustrated in figure~\ref{het_fig2},
particularly around $X=90$ and $X=140$ (where the solution is close to
$P_+$), and at $X=120$ where the solution is close to $P_-$. For the
parameter values of the figure, the pure mode equilibria have
amplitude $|B|=0.2$. Nearby spatial locations are often still
well-correlated in time, see figure~\ref{het_fig}.


\section{Stable travelling waves and complex spatiotemporal dynamics}

\label{stable_tw_sec}

\begin{figure}[!h]
\begin{center}
\includegraphics[width=12.0cm]{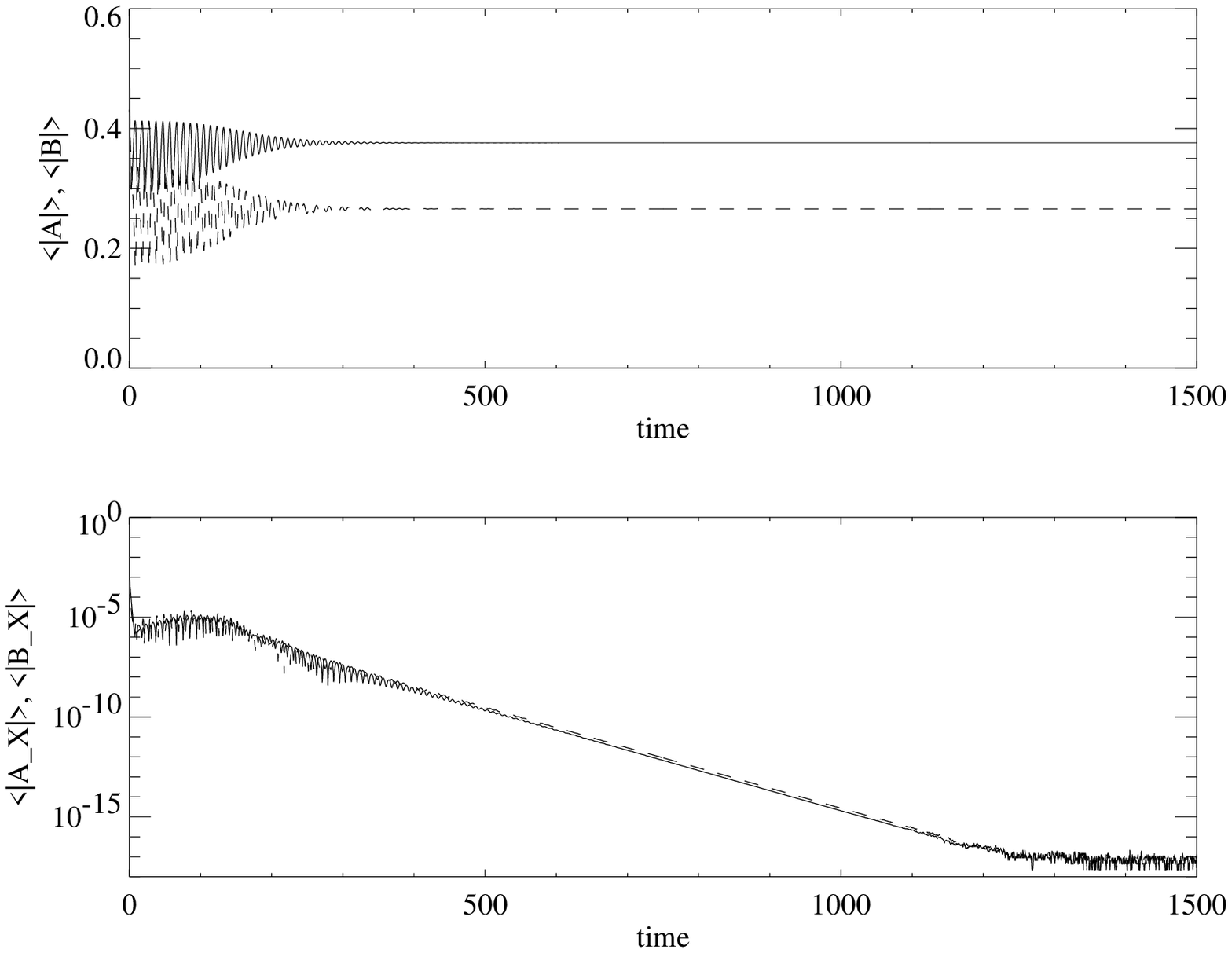}
\includegraphics[width=12.0cm]{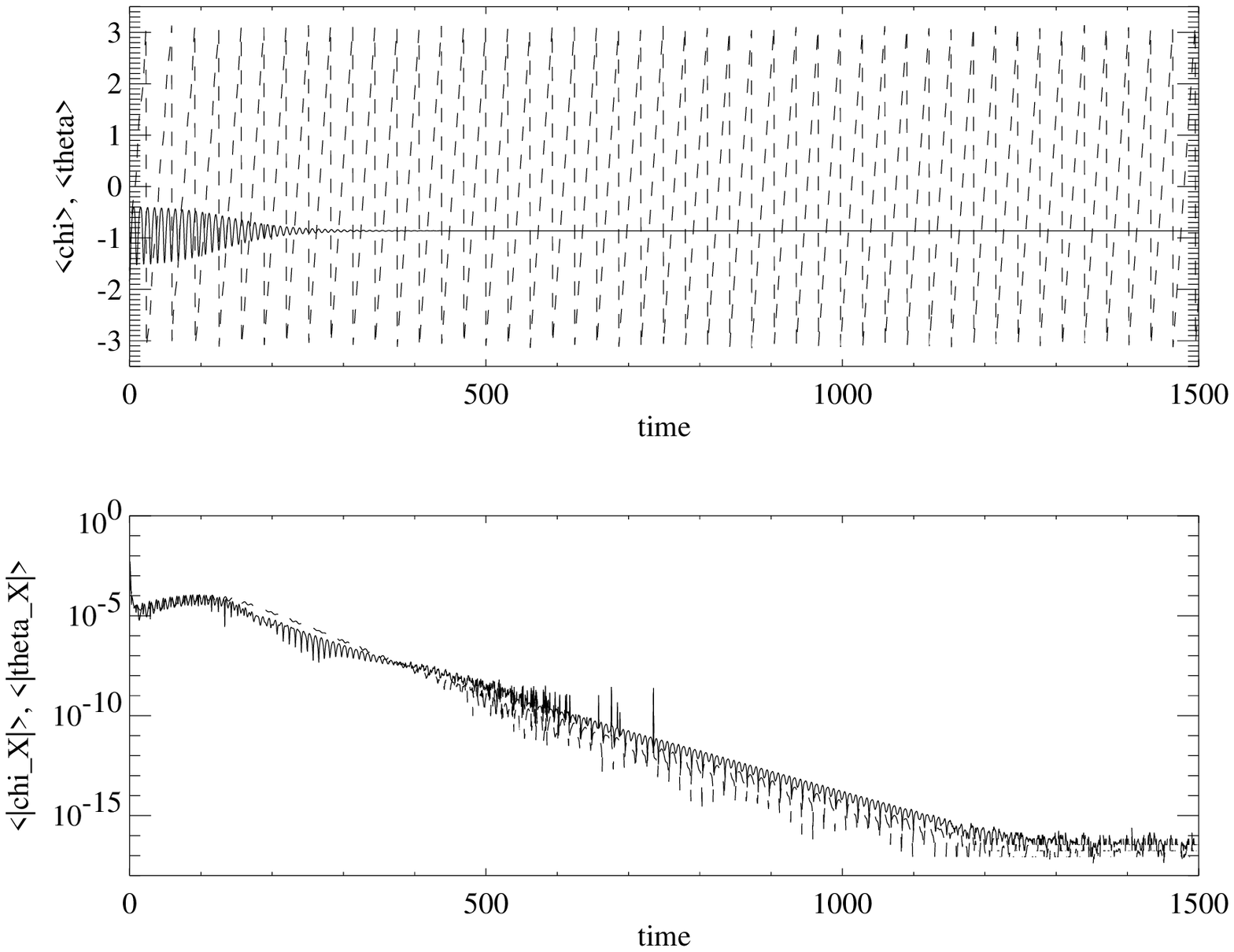}
\caption{Temporal evolution of the spatial averages of the amplitude
and phase variables at $\mu_1=0.15$, $\mu_2=0.7$, $L=30$; all other
coefficients are as in figure~\ref{mu1mu2_fig}. The four panels show
spatial averages of: (a) $|A|$ (solid
line) and $|B|$ (dashed line); (b) $|A_X|$ (solid) $|B_X|$ (dashed);
(c) $\newchi$ (solid), $\theta$ (dashed); (d) $|\newchi_X|$ (solid),
$|\theta_X|$ (dashed). After an oscillatory
transient the solution is attracted towards a spatially-periodic TW
state where $\theta$ (and $\phi$, not shown) increase
linearly with time while $|A|$, $|B|$ and $\newchi$ remain constant.}
\label{stable_tw_fig}
\end{center}
\end{figure}

\begin{figure}[!h]
\begin{center}
\includegraphics[width=12.0cm]{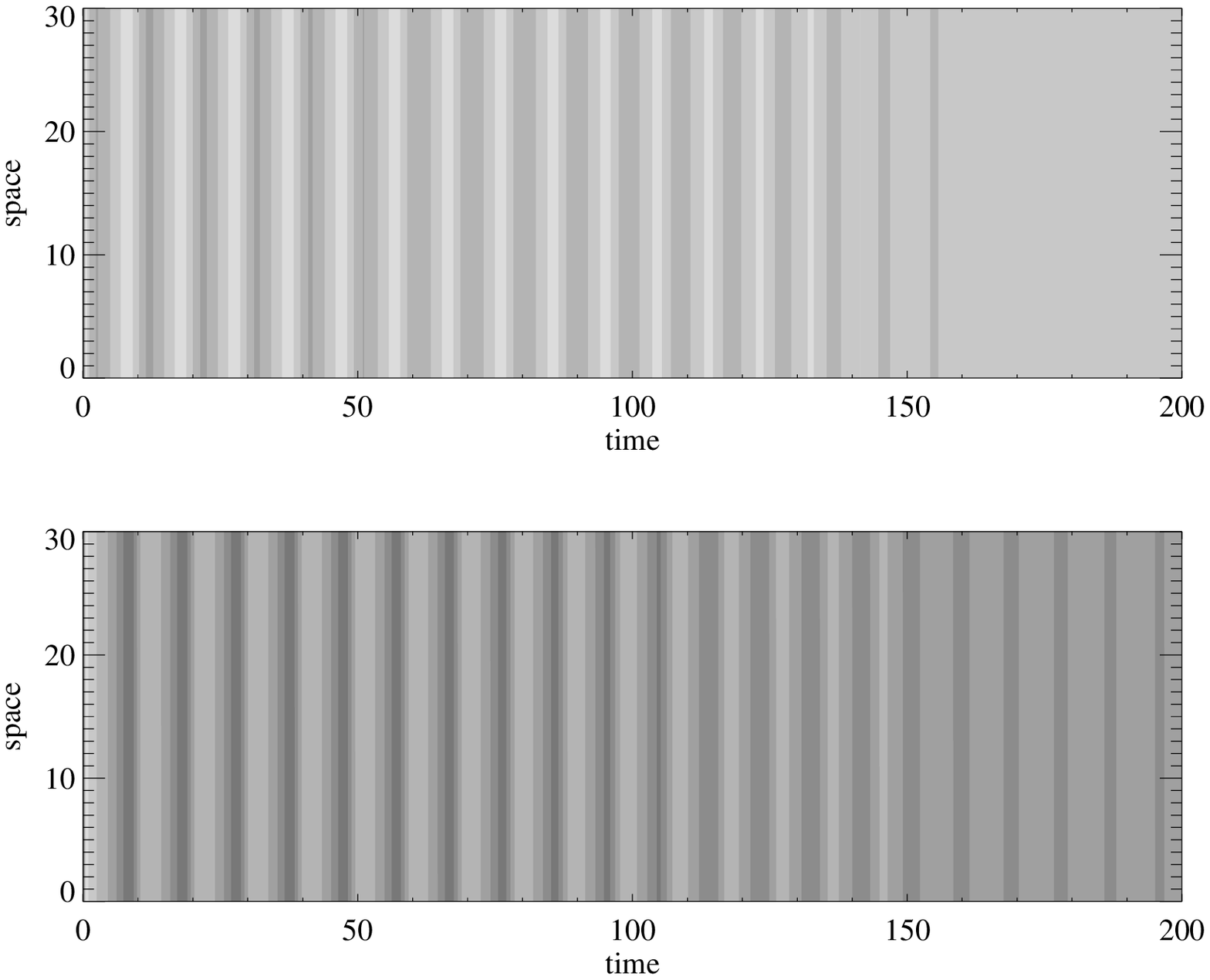}
\caption{The spatio-temporal evolution of the amplitudes $|A|$ (upper
panel) and $|B|$ (lower panel) for the parameter values of
figure~\ref{stable_tw_fig}, showing convergence to TW.}
\label{stable_tw_st_fig}
\end{center}
\end{figure}

\begin{figure}[!h]
\begin{center}
\includegraphics[width=12.0cm]{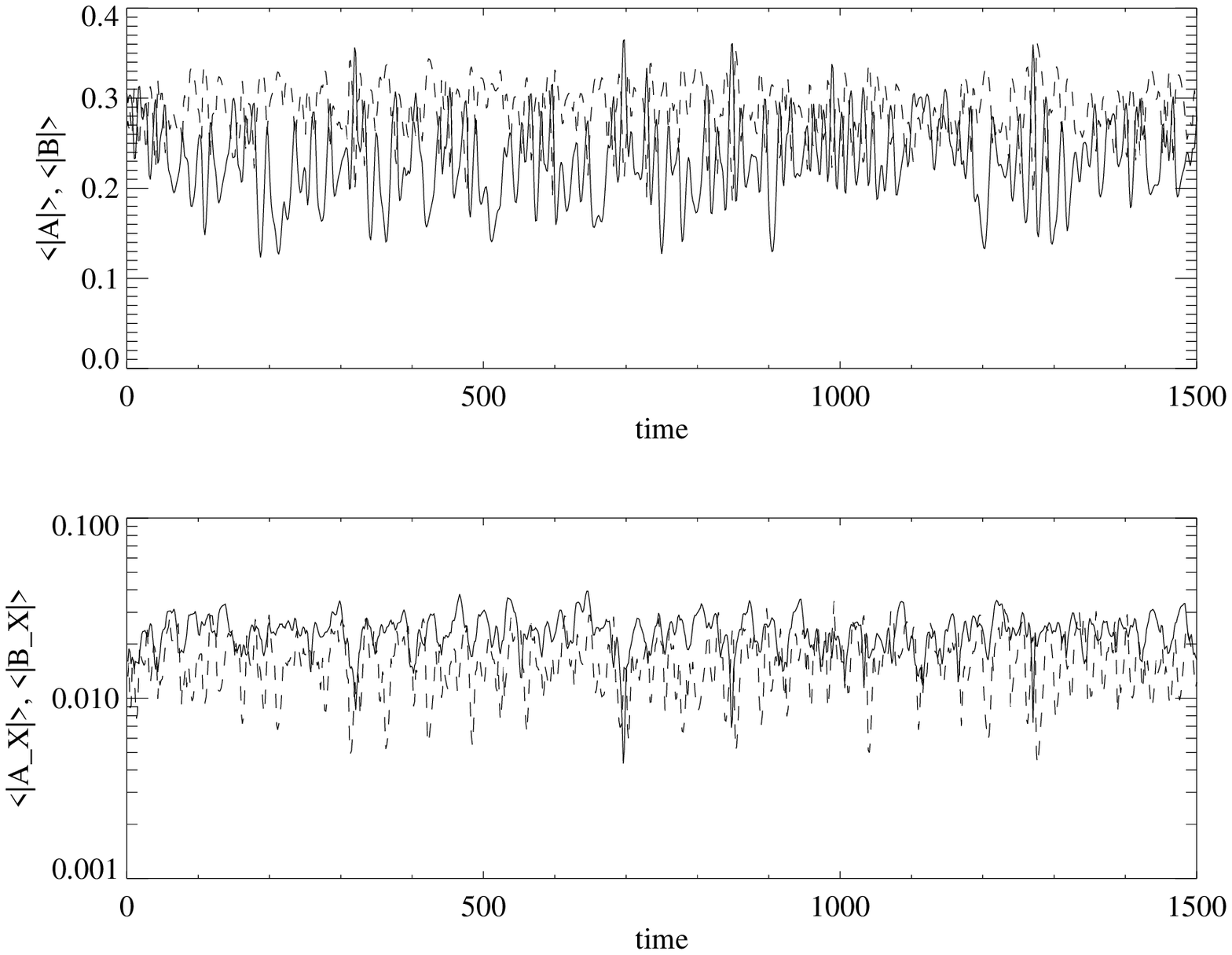}
\includegraphics[width=12.0cm]{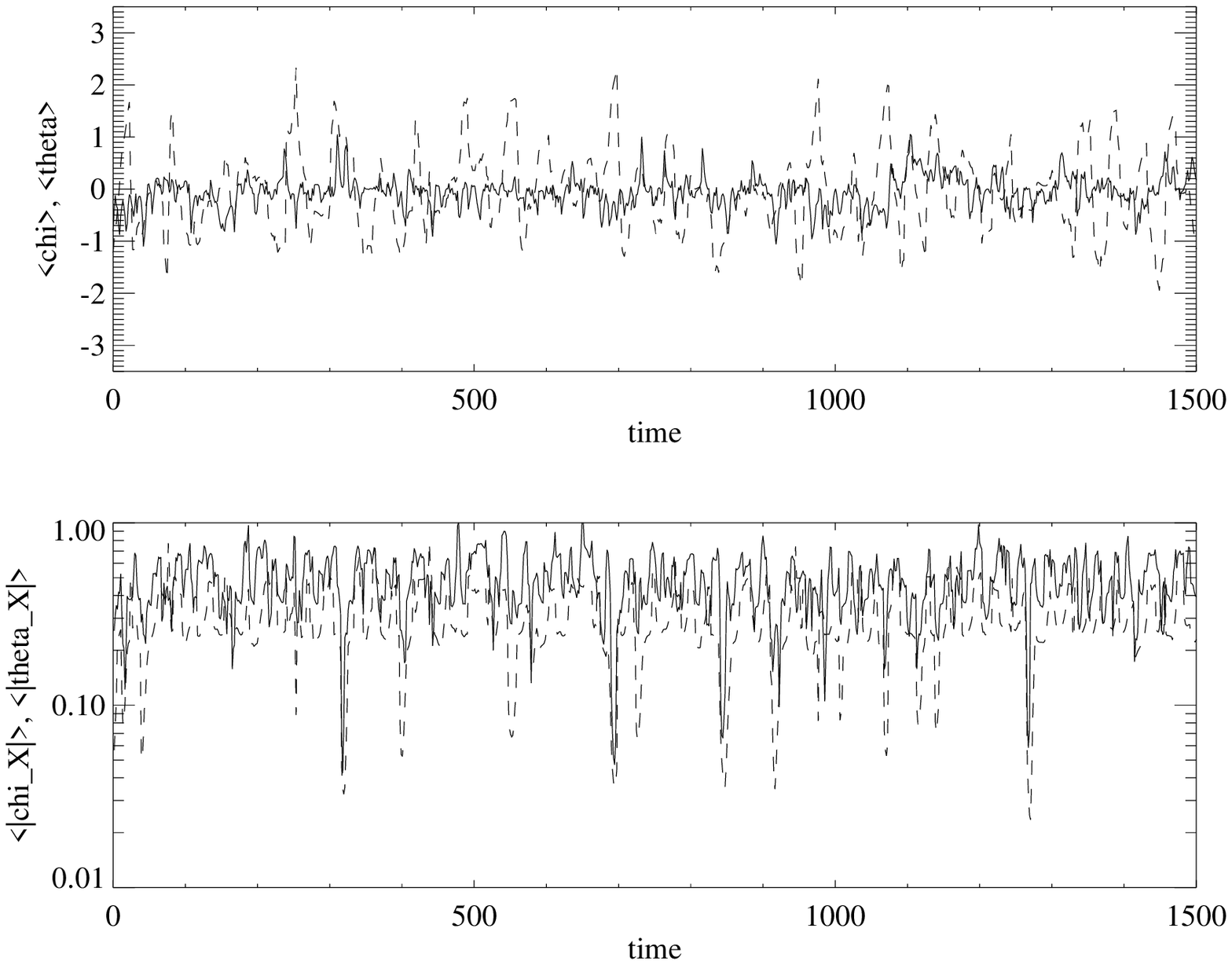}
\caption{Temporal evolution of a persistent complex spatiotemporal
state at $\mu_1=0.15$, $\mu_2=0.7$, $L=30$; all other coefficients
are as in figure~\ref{mu1mu2_fig}. The four panels show spatial
averages of: (a) $|A|$ (solid
line) and $|B|$ (dashed line); (b) $|A_X|$ (solid) $|B_X|$ (dashed);
(c) $\newchi$ (solid), $\theta$ (dashed); (d) $|\newchi_X|$ (solid),
$|\theta_X|$ (dashed). This solution coexists stably with the TW state
shown in figure~\ref{stable_tw_fig}.}
\label{stable_stc_fig}
\end{center}
\end{figure}

\begin{figure}[!h]
\begin{center}
\includegraphics[width=12.0cm]{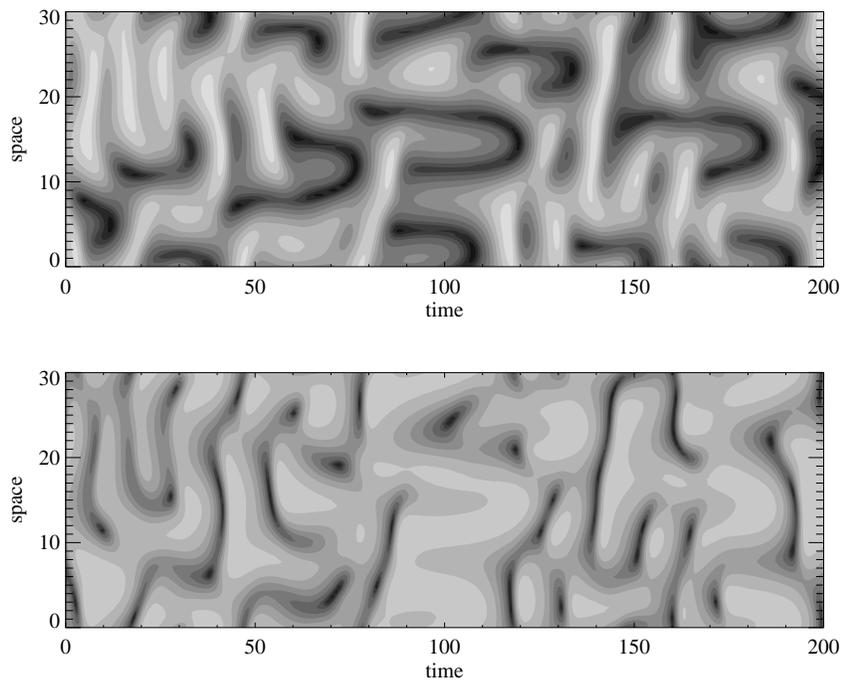}
\caption{Spatio-temporal evolution of the amplitudes $|A|$ (upper
panel) and $|B|$ (lower panel) for the parameter values of
figure~\ref{stable_stc_fig}, showing complicated spatiotemporal
dynamics.}
\label{stable_stc_st_fig}
\end{center}
\end{figure}

At large positive $\mu_1$ and $\mu_2$ the TW states are restabilised
after a Hopf bifurcation involving a branch of modulated waves
generated by the global bifurcation in which the heteroclinic cycle
loses stability. Surprisingly, for the spatially-extended system, these
spatially-homogeneous TW states are also stable for large enough
$\mu_1$ and $\mu_2$, as shown in figures~\ref{stable_tw_fig} and~\ref{stable_tw_st_fig}.
The stable TW coexist with stable complicated spatiotemporal dynamics,
illustrated in figures~\ref{stable_stc_fig}
and~\ref{stable_stc_st_fig}. The latter state might 
be expected to be a more generic solution in this region of the
$(\mu_1,\mu_2)$ plane, although no systematic study of the relative
sizes of the basins of attraction has been performed.

\clearpage


\section{Discussion and conclusions}

\label{sec:conclusions}

In this paper we have examined the robustness of the results of Part 1
on the dynamics near the $1:2$ strong spatial resonance, to
modulational instabilities which might be expected to play a role when
the spatial resonance is close to, but not exactly, $1:2$. We have
derived a pair of coupled amplitude equations using symmetry
arguments; these amplitude equations are PDEs rather than ODEs as is
the case for exact resonance. 

We have examined the stability of the spatially-periodic solutions
found in Part 1, looking for instabilities to finite-wavelength and
long-wavelength perturbations. In general, we find that the
results of Part 1 still apply when $\mu_1$ and $\mu_2$ are large
(equivalently when the mismatch $q$ is small), but that new
instabilities are present near the codimension-two point
$\mu_1=\mu_2=0$. Since these lead, in general, to variations with a
horizontal wavenumber $\O(\epsilon q)$, solutions still resemble those
of the non-modulated problem over short lengthscales. 

This paper is primarily concerned with identifying the several
codimension-two points in the parameter space which link modulational
instabilities and amplitude or phase instabilities which are present
in the ODEs discussed in Part 1. We have presented reduced
equations that describe the dynamics near these various points. The
form of these reduced equations (which are still PDEs) may be deduced
from symmetry arguments, and can also be derived through perturbation
expansions. This analysis
enables us to understand qualitative features of the dynamics;
although analytic solutions are still extremely difficult to derive
for the reduced equations, this process of deriving `normal forms'
contributes greatly to an understanding of the interaction of these
various instabilities.

Our three main results are firstly, that the pure mode $P$ and the
mixed-mode $M_+$ may undergo steady-state instabilities to
finite-wavelength modes with a maximum critical wavenumber $\ell_c=q$
(for $P$) and $\ell_c=2q$ (for $M_+$).
This confirms our intuition; it is natural to think that there
are only two important lengthscales in the problem - the $O(1)$
lengthscale of the spatially-periodic patterns and long lengthscales
$>O(1/\epsilon q)$ introduced by the deviation from exact
resonance. $M_+$ may also undergo an oscillatory bifurcation to a
non-zero wavenumber mode, and numerical investigations suggest this
wavenumber is also $O(q)$.
Secondly, the reduction of the governing PDEs to a `normal form' and
subsequent analysis of these reduced equations for the codimension-two
phase instability at the point $\X$ is novel and is presented in detail.
Thirdly, we have considered the fate of the structurally stable
heteroclinic cycle present in the ODE problem. Although the equilibria
on the cycle, and hence the cycle itself, must be unstable to
long-wavelength perturbations, the
dynamics for small $q$ result in a stable periodic orbit that lies
close to the cycle and whose period can be estimated well
analytically. At larger $q$ the heteroclinic cycle still organises the
spatial dynamics over short distances, although solutions are
spatiotemporally complex as might in general be expected.

There is, of course, much more work that could be done on this
problem. Two obvious directions of interest are the fate of the more
complicated heteroclinic cycles examined by Porter \& Knobloch
\cite{PK00} to modulational perturbations, and the behaviour for
asymptotically large $q$ (within this scaling ansatz); we have
concentrated on behaviour for $q \leq 1$, where the results may be
most easily interpreted with reference to previous work.

\begin{ack}
We are indebted to Chris Jones and Philippe Metzener for various
stimulating discussions on aspects of this problem. We are grateful to
the authors of \cite{HRS03} for communicating a copy of their
results prior to publication. We would also like
to thank Jennifer Siggers for many helpful comments and technical
assistance. CMP and JHPD are grateful for financial support from
Trinity College, Cambridge.
\end{ack}


\section*{Appendix}

\label{appendix1}

In this appendix we give the coefficients in the Ginzburg--Landau
equation governing the instability of the pure mode solution $P_+$ 
along the interior of the line $\Origin\L$, discussed in section~\ref{sec:pure_mode}.

The $P_+$ equilibrium is given by $A=0$, $B=\sqrt{\mu_2/a_2}=B_0$. We
start from the ansatz
\ba
A(X,T) & = & \epsilon \left[ \alpha(\xi,\tau) B_0 \e^{\i\ell X} + \bar{\alpha}(\xi,\tau)
2q(q+\ell) \e^{-\i\ell X} \right] \nn \\
& & + \epsilon^2 \left[ \gamma_0(\xi,\tau)+\gamma_1(\xi,\tau)
\e^{2\i\ell X} + \gamma_2(\xi,\tau) \e^{-2\i\ell X} \right] + O(\epsilon^3), \label{app:eqn1} \\
B(X,T) & = & B_0 + \epsilon^2 \left[ \beta_0(\xi,\tau) +
\beta_1(\xi,\tau)\e^{2\i\ell X} + \beta_2(\xi,\tau) \e^{-2 \i\ell X} \right] +
O(\epsilon^3), \label{app:eqn2}
\ea
where $\xi=\epsilon X$ is a new, longer, lengthscale and
$\tau=\epsilon^2 T$ is a new slower timescale. After a multiple-scales
expansion, substituting this ansatz into the PDEs~(\ref{eqn1}) -
(\ref{eqn2}), the perturbation
amplitudes $\beta_0,\beta_1,\beta_2,\gamma_0,\gamma_1,\gamma_2$ can
be expressed in terms of $\alpha$. The usual solvability condition at
$O(\epsilon^2)$ is satisfied identically, and at $O(\epsilon^3)$ the
solvability condition yields the Ginzburg--Landau amplitude equation
for $\alpha$:
\ba
\hat{c}_1 \alpha_\tau & = & \hat{c}_1 \alpha_{\xi \xi} - \hat{c}_2
\alpha |\alpha|^2, \nn
\ea
where the real coefficients $\hat{c}_1$ and $\hat{c}_2$ are
\ba
\hat{c}_1 & = & B_0^2 + 4q^2(q+\ell)^2, \nn \\
& & \nn \\
\hat{c}_2 & = & a_1 [B_0^4 + 16 q^4(q+\ell)^4 + 16 B_0^2
q^2(q+\ell)^2] \nn \\
& & + \hat{c}_3 B_0 [2b_1B_0^2 + 8q^2(q+\ell)^2b_1 - 4q(q+\ell)] \nn \\
& & + 4b_1B_0^2q(q+\ell)(\hat{c}_4+\hat{c}_5) - B_0^2\hat{c}_4 - 4q^2(q+\ell)^2\hat{c}_5, \nn \\
& & \nn \\
\hat{c}_3 & = & -[B_0^2b_2 + 4b_2 q^2(q+\ell)^2 + 4q(q+\ell)]/(2a_2 B_0), \nn \\
& & \nn \\
\hat{c}_4 & = & \frac{2 B_0^2 q(q+\ell)[a_2 q(q+\ell) - 4 b_2 c\ell^2]
  - B_0^2(a_2 B_0^2 + 4c\ell^2)}{8c\ell^2(a_2 B_0^2 + 2c\ell^2)}, \nn \\
& & \nn \\
\hat{c}_5 & = & \frac{a_2 B_0^4 -
2q(q+\ell)[q(q+\ell)(a_2 B_0^2 + c\ell^2)+4b_2 B_0^2 c \ell^2
]}{8c\ell^2(a_2 B_0^2 + 2c\ell^2)}. \nn
\ea
Hence $\hat{c}_2>0$ when $a_1$ is sufficiently positive. In the
special case $\hat{c}_3=\hat{c}_4=\hat{c}_5=0$, the
expression for $\hat{c}_2$ agrees with that derived by Coullet et
al. \cite{CRV86}. In fact, the $O(\epsilon^2)$ amplitudes
$\gamma_0,\gamma_1,\gamma_2$ in~(\ref{app:eqn1}) do not contribute to the
complexity of the expression for $\hat{c}_2$; the complexity is
derived from the third-order contributions from the $\dot{B}$
equation.

\end{document}